\documentclass[a4paper]{article}
\usepackage{RR}
\usepackage{hyperref}
\usepackage{graphicx}
\usepackage{float}
\usepackage{manfnt}
\usepackage{fancybox}
\usepackage{tabularx}
\usepackage{color}
\usepackage{multirow}
\RRdate{March 2010}
\RRauthor{
Anne-C\'ecile Orgerie
  \and
Laurent Lef\`evre
}
\authorhead{Orgerie \& Lef\`evre}
\RRtitle{Une ann\'ee dans la vie d'un syst\`eme distribu\'e exp\'erimental \`a grande \'echelle : la plate-forme Grid'5000 en 2008}
\RRetitle{A year in the life of a large scale experimental distributed system: the Grid'5000 platform in  2008}
\titlehead{A year in the life of the Grid'5000 platform (2008)}
\RRresume{Ce rapport pr\'esente les r\'esultats d'usage de la plateforme exp\'erimentale Grid'5000 pendant l'ann\'ee 2008. L'usage des principaux sites opérationnels de Grid'5000 (Bordeaux, Lille, Lyon, Nancy, Orsay, Rennes, Sophia-Antipolis, Toulouse) est presenté et analysé.}
\RRabstract{This report presents the usage results of Grid'5000 over year 2008. Usage of the main operationnal Grid'5000 sites (Bordeaux, Lille, Lyon, Nancy, Orsay, Rennes, Sophia-Antipolis, Toulouse) is presented and analyzed.}
\RRmotcle{Système distribué à grande échelle, usage, Grid'5000}
\RRkeyword{Large scale distributed system, usage, Grid'5000}
\RRprojets{Reso}
\RRNo{7481}
\RRtheme{\THNum} 
\URRhoneAlpes

\definecolor{rouge}{rgb}{0.64,0.06,0.06}

\hypersetup{
     backref=true,    
     pagebackref=true,
     hyperindex=true, 
     colorlinks=true, 
     breaklinks=true, 
     urlcolor= blue, 
     linkcolor= blue, 
     citecolor=rouge,
     bookmarks=true, 
     bookmarksopen=true,            
}

\begin{document}
\makeRR   

\graphicspath{{bordeaux/}{grenoble/}{rennes/}{toulouse/}{sophia/}{lyon/}{nancy/}{lille/}{orsay/}{grid/}}

\tableofcontents
\newpage

\section{Introduction}

Some previous work on operational Grids~\cite{IDE06} show that grids are not utilized at their full capacity. We focus on the utilization of a large-scale experimental distributed system by relying on the case study of Grid'5000\cite{G5K}\footnote{Some experiments of this article were performed on the Grid'5000 platform, an initiative from the French Ministry of Research through the ACI GRID incentive action, INRIA, CNRS and RENATER and other contributing partners (http://www.grid5000.fr). This research is supported by the GREEN-NET INRIA Cooperative Research Action: \url{http://www.ens-lyon.fr/LIP/RESO/Projects/GREEN-NET/}}, a French experimental Grid. We collected and analyzed the logs of the Grid'5000 platform for a full year of usage (2008). We present the main statistics observed for each site and some specific and representative computing nodes. A special analysis is dedicated to the grid usage of this experimental platform. This usage analysis has been used in some works on energy-efficient experimental distributed infrastructures~\cite{OLG08a,OLG08b}. A similar analysis has been done for 2007 logs of Grid'5000 and can been found in~\cite{rapport2007}.

This paper briefly presents the Grid'5000 experimental platform and the followed methodology in Section~\ref{g5k}. The usage statistics are then presented for each Grid'5000 site (Section~\ref{results}).

\section{Definitions and Methodology}
\label{g5k}
\subsection{Grid'5000: a Large-Scale Experimental Distributed System}
\begin{figure}[H]
  \centering
    \includegraphics[width=5cm]{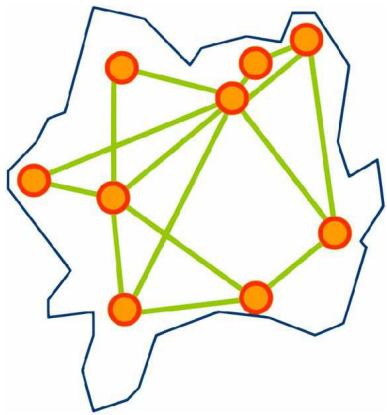}
    \caption{The Grid'5000 map}
    \label{platform}
\end{figure}

The Grid'5000 platform is an experimental testbed dedicated for research in computer science, made up of more than 3400~processors geographically distributed on 9 sites in France (see Fig.~\ref{platform}). This platform can be defined as a highly reconfigurable, controllable, and monitorable experimental large-scale distributed system.

The utilization of Grid'5000 can be very specific. Each user can reserve (in advance) some computing nodes. During its reservation time, the user can be root on his reserved nodes and he can deploy his own system images, collect data, launch applications, reboot the nodes, and so on. The nodes are entirely dedicated to the user during his reservation.

\subsection{Experiment Methodology}

The user gives the resource manager (at least) a start time, a
duration, and the number of required resources. These characteristics
define the job (a reservation). The resource manager is in charge of
the job's acceptance. It verifies if that job is compatible with
previously accepted jobs and, if it is, it gives a job id
to the user.

There are three different type of jobs:
\begin{itemize}
\item \textit{deploy}: the user deploys its own environment,
\item \textit{default}: the user uses the default environment.
\item \textit{besteffort}: they get scheduled on processes when no other job use them and these jobs are killed when a regular (deploy or default) job recently submitted needs the nodes used by a besteffort job.
\end{itemize}

When a resource is not available for the user, it can be in different states:
\begin{itemize}
\item {\bf dead}: the resource is down (due to a component failure for example); 
\item {\bf suspected}: the resource does not work properly;
\item {\bf absent}: the resource is not available for the user (not physically present).
\end{itemize}

Moreover, the platform has changed between the beginning and the end of the measurements. For that reason, we present the results as percentages of the platform's capacity at the time of measurement (100\% at a given time does not represent the same number of resources as 100\% at another time).

In order to obtain the utilization traces, we have used a history function provided by OAR\footnote{OAR is a resource manager (or batch scheduler) for large clusters.(\url{http://oar.imag.fr/index.html})}~\cite{oar} called \textit{oarstat}. This function provides the user with all the events that occur during a time period. An event can be a job or a problem on a resource (it is dead, absent or suspected) for a given period. Our goal is to obtain a detailed overview of the usage of an experimental grid.

\section{Results per Grid'5000 site}
\label{results}
For each Grid'5000 site, we provide several values and figures which represent the global usage of the site. All the times are given in seconds. A resource is a core and a job is actually a reservation.

We have split the proposed statistics in three categories:
\begin{itemize}
\item the ``platform and resources'' part which lists the available number of resources and some statistics on the resources' states. We also compute the real work time of the Grid'5000 site. The ``real'' percentage of work time is calculated without taking into account dead time or absent time (i.e. work time over total time minus dead time and absent time for all the resources). 
\item the ``jobs'' part proposes some statistics on the number of submitted jobs (reservations), the mean time and mean number of resources per job, and the maximal duration of a job on a site. We also list the number of reservations used for deploying system images ({\it deploy jobs}).
\item the ``users'' section contains the number of users and the impact of users from other Grid'5000 sites. The percentage of users coming from other sites does not take into account any consideration of proportionality in terms of jobs. It is just the number of users coming from other sites over the total number of users (each user is counted once).
\end{itemize}

For each site, we provide four diagrams. The first one represents the weekly repartition in time of the resources between the different states: in red when some cores are dead ({\it Dead} state), in orange when they are suspected ({\it Suspected} state), in yellow when they are absent ({\it Absent} state), in green when they are working (a job is running) ({\it Work} state), and in white when they are unoccupied (no job, no other state) ({\it Idle} state).

The two other diagrams present the weekly time repartition of two particular resources: the median and the maximal resources. The "maximal resource" is the resource which has the maximal work time among the resources which are present for the whole 2008 year. The "median resource" is a resource which is present for the whole 2008 year and is the nearest to the median value of cumulative work over the experiment's duration.

\subsection{Usage of Grid'5000's Bordeaux site in 2008}
\begin{itemize}
\item Platform and resources:
\begin{itemize}

\item Maximal number of resources (cores): 650
\item Mean time spent in each state for all the resources, in percentage:
\begin{itemize}
\item Dead: 3.48\%
\item Suspected: 5.53\%
\item Absent: 4.06\%
\item Work: 49.19\%
\end{itemize}
\item Real percentage of work time (without taking into account the time when the resources are dead or absent): 53.20\%
\end{itemize}

\item Jobs:
\begin{itemize}
\item Number of jobs (reservations): 356222
\item Mean time of a job: 2473.38 (41 minutes and 13 seconds)
\item Maximal duration:  794428 s. (9 days 4 hours 40 minutes and 28 seconds) for job number 407783
\item Mean number of resources (cores) per job: 7.44
\item Percentage of deploy jobs: 0.97\%
\item Percentage of time spent in deploy jobs compared to the work time: 22.32\%
\item Percentage of jobs coming from other sites: 83.28\%
\item Percentage of besteffort jobs: 58.03\%
\item Percentage of time spent in besteffort jobs compared to the work time: 34.90\%
\end{itemize}

\item Users:
\begin{itemize}
\item Number of users: 223
\item Percentage of users coming from other sites: 87\%
\end{itemize}
\end{itemize}

\begin{figure}[H]
  \centering
    \includegraphics[width=14cm]{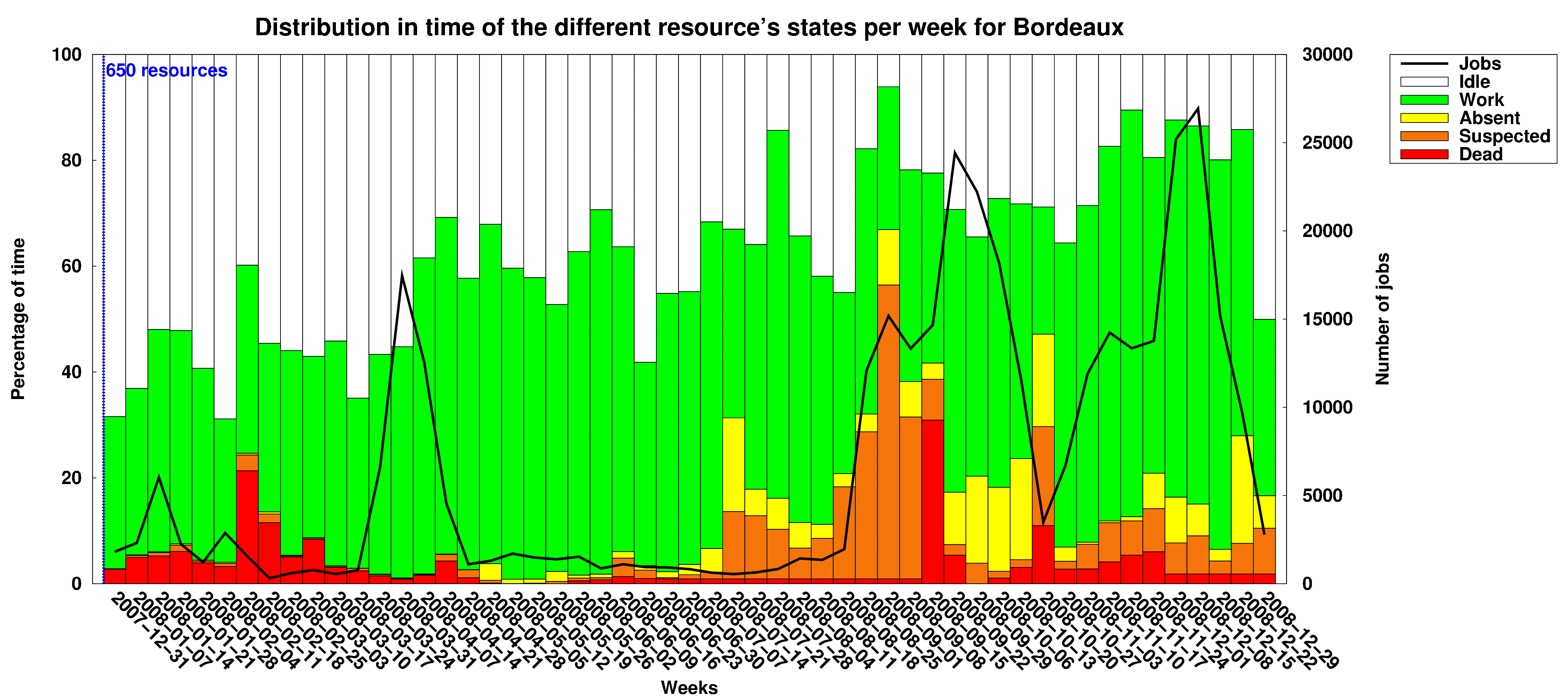}
    \caption{Global diagram with dead time for Grid'5000's Bordeaux site}
    \label{bdx1}
\end{figure}

\begin{figure}[H]
  \centering
   \includegraphics[width=14cm]{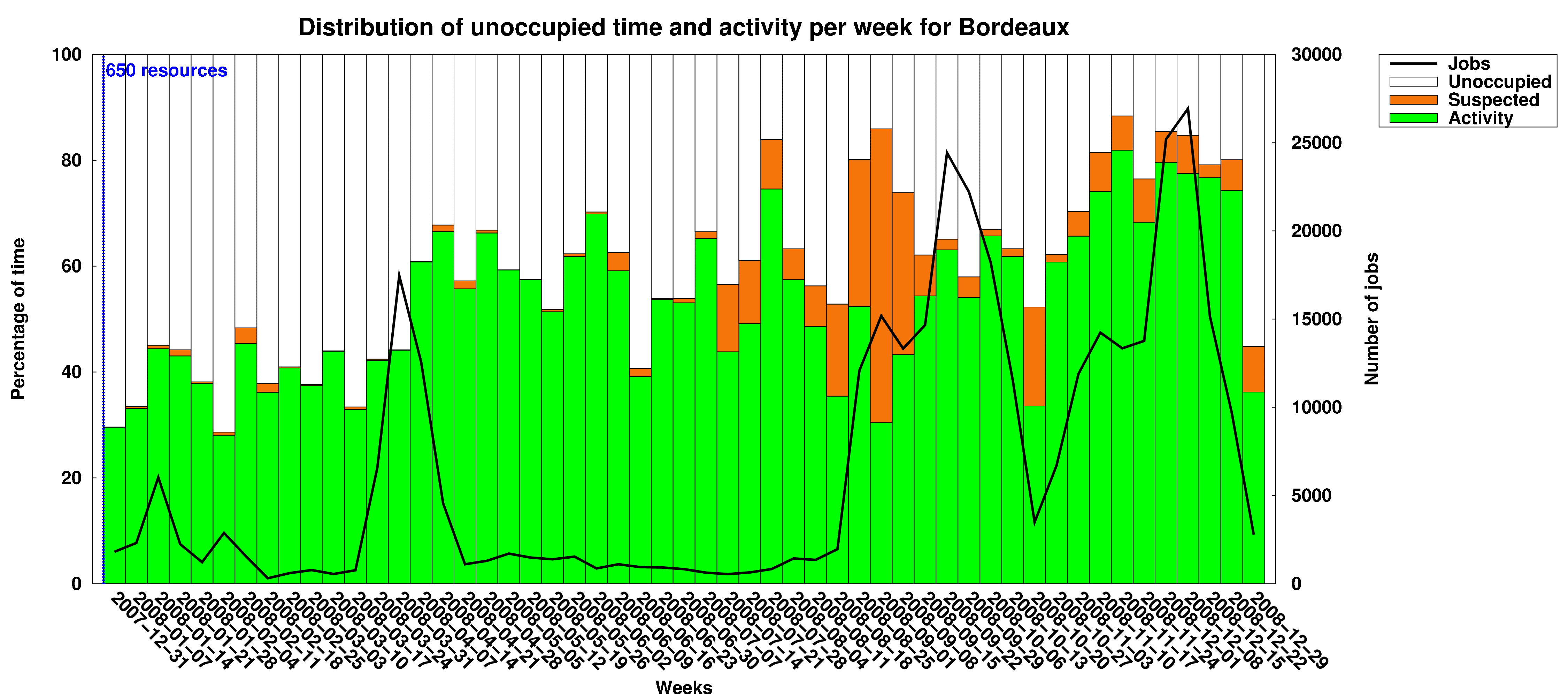}
   \caption{Global diagram without dead time for Grid'5000's Bordeaux site}
    \label{bdx2}
\end{figure}


\begin{figure}[H]
  \centering
    \includegraphics[width=14cm]{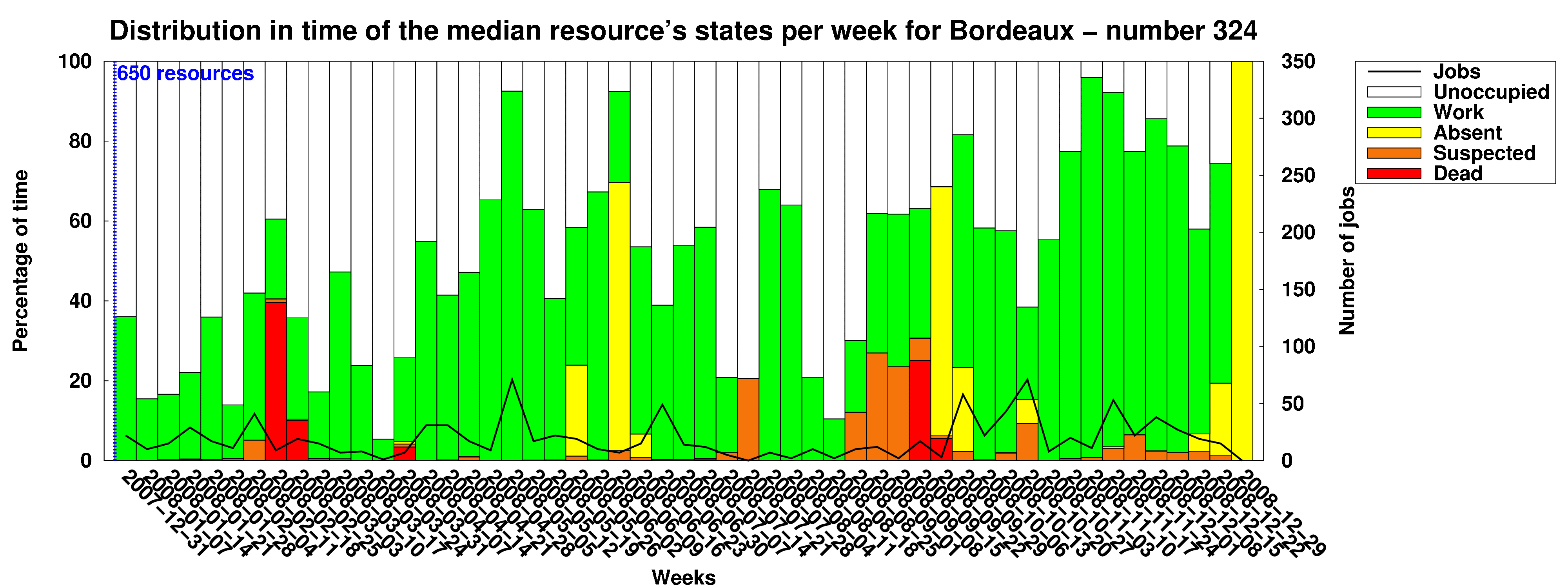}
    \caption{Median resource diagram for Grid'5000's Bordeaux site}
    \label{bdxmed}
\end{figure}

\begin{figure}[H]
  \centering
    \includegraphics[width=14cm]{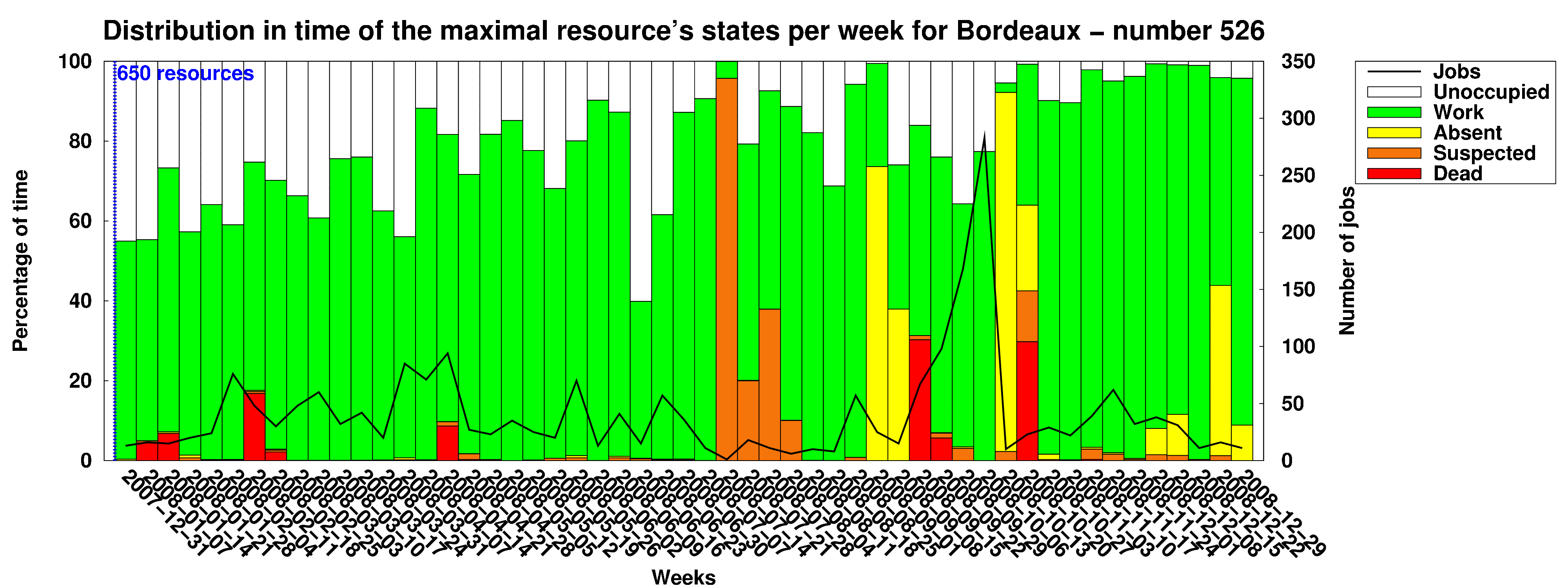}
    \caption{Maximal resource diagram for Grid'5000's Bordeaux site}
    \label{bdxmax}
\end{figure}

$\textdbend$ We can see on these graphs that the site load is not related to the number of jobs. Indeed, between May and August, there are few jobs but the percentage of activity is high (Figure~\ref{bdx1}). Bordeaux is the less used Grid'5000 site for 2008 in terms of real activity. However, its activity has increased compared to the previous year~\cite{rapport2007}.

\subsection{Usage of Grid'5000's Lille site in 2008}
\begin{itemize}
\item Platform and resources:
\begin{itemize}

\item Maximal number of resources (cores): 618
\item Mean time spent in each state for all the resources, in percentage:
\begin{itemize}
\item Dead: 16.21\%
\item Suspected: 4.67\%
\item Absent: 8.61\%
\item Work: 54.80\%
\end{itemize}
\item Real percentage of work time (without taking into account the time when the resources are dead or absent): 72.89\%
\end{itemize}

\item Jobs:
\begin{itemize}
\item Number of jobs (reservations): 344538
\item Mean time of a job: 3154.58 s. (52 minutes and 34 seconds)
\item Maximal duration:  756077 s. (8 days 18 hours 1 minute and 17 seconds) for job number 693546
\item Mean number of resources (cores) per job: 8.11
\item Percentage of deploy jobs:  0.61\%
\item Percentage of time spent in deploy jobs compared to the work time: 7.24\%
\item Percentage of jobs coming from other sites: 38.97\%
\item Percentage of besteffort jobs: 19.01\%
\item Percentage of time spent in besteffort jobs compared to the work time: 37.88\%
\end{itemize}

\item Users:
\begin{itemize}
\item Number of users: 198
\item Percentage of users coming from other sites: 65.66\%
\end{itemize}
\end{itemize}

\begin{figure}[H]
  \centering
    \includegraphics[width=14cm]{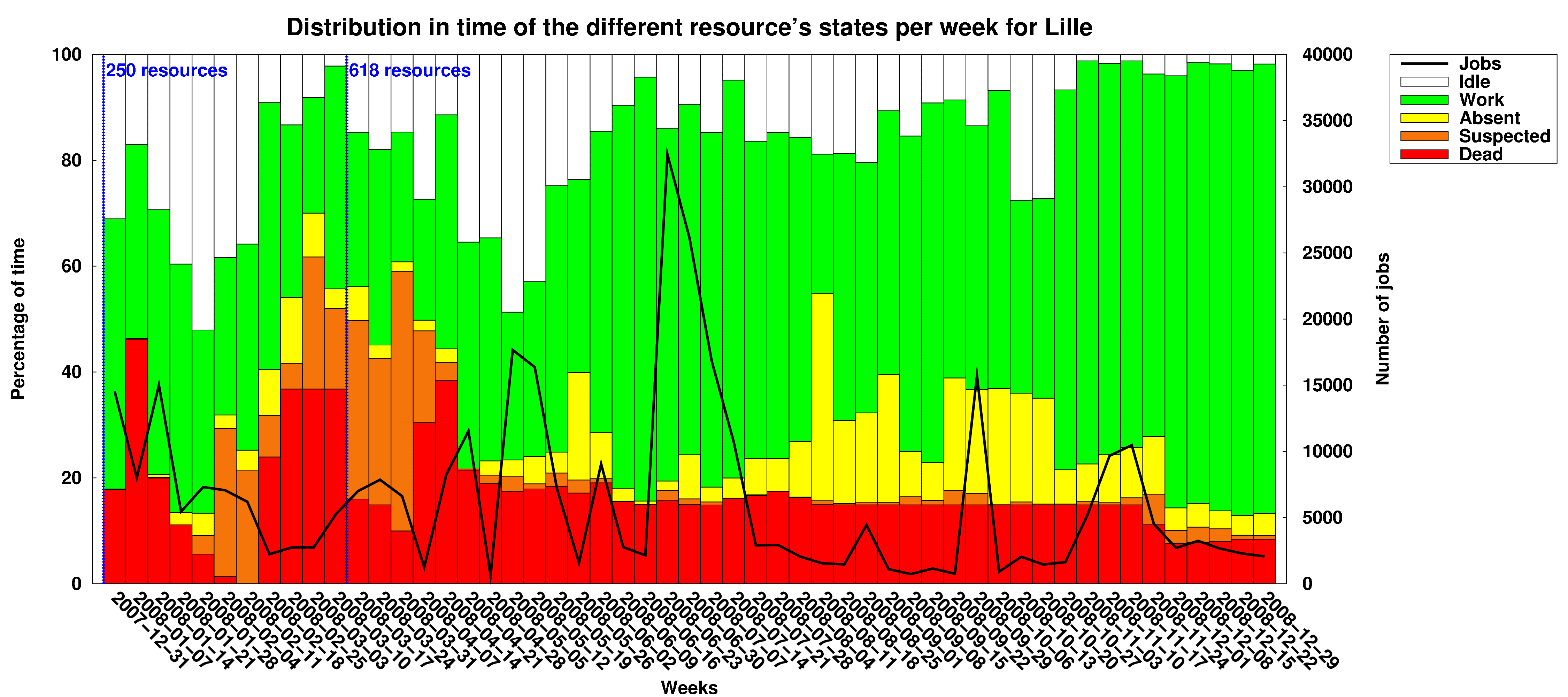}
    \caption{Global diagram with dead time for Grid'5000's Lille site}
    \label{lil1}
\end{figure}

\begin{figure}[H]
  \centering
   \includegraphics[width=14cm]{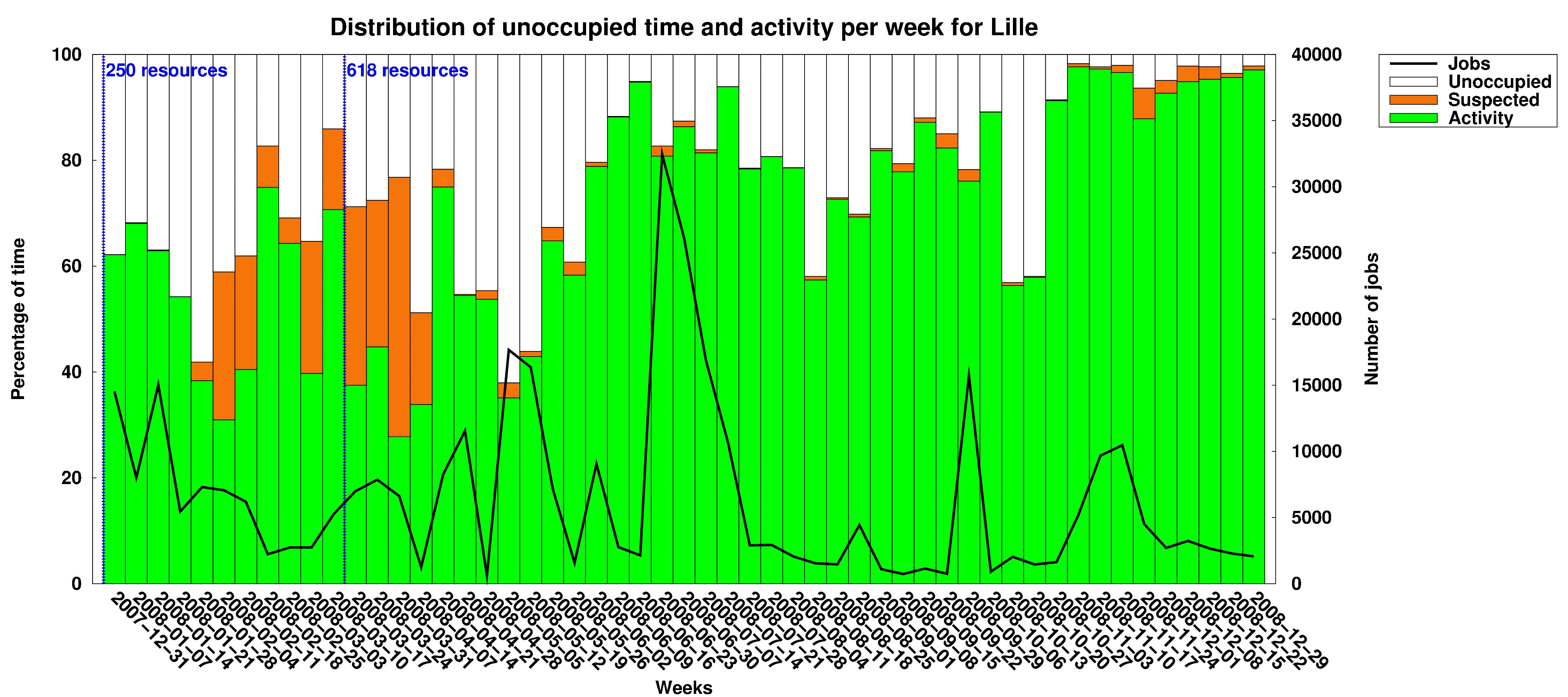}
   \caption{Global diagram without dead time for Grid'5000's Lille site}
    \label{lil2}
\end{figure}


\begin{figure}[H]
  \centering
    \includegraphics[width=14cm]{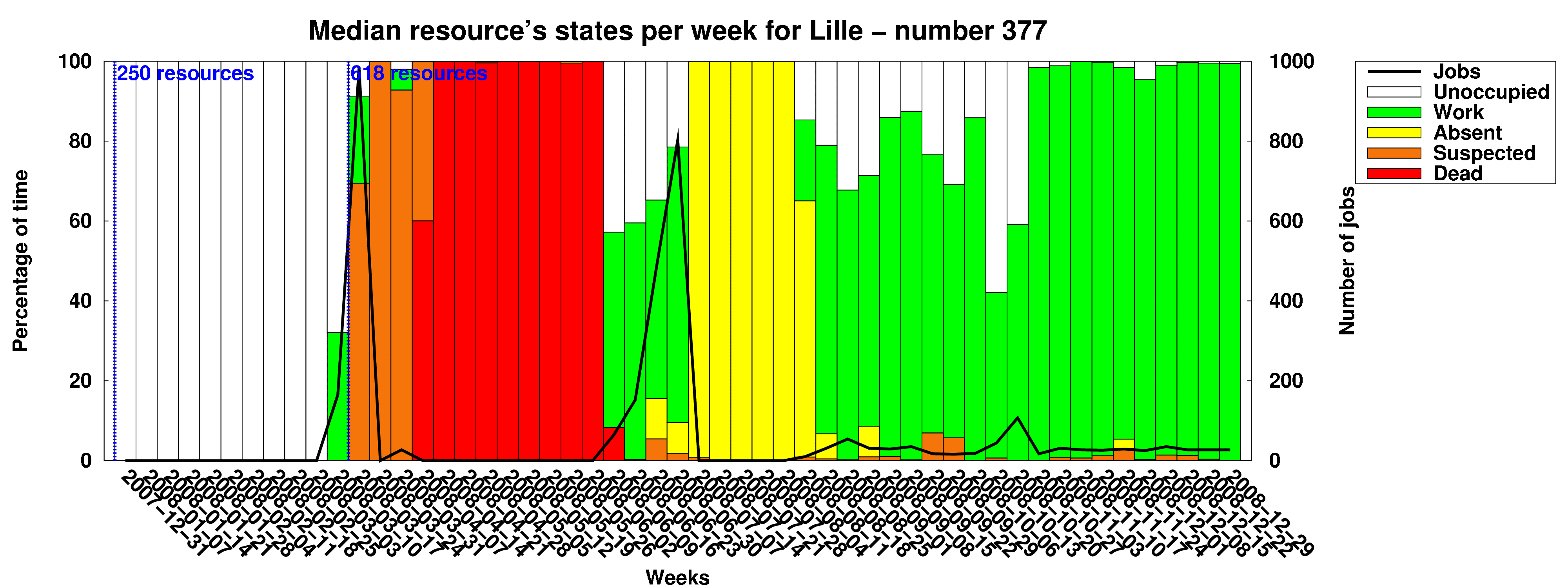}
    \caption{Median resource diagram for Grid'5000's Lille site}
    \label{lilmed}
\end{figure}

\begin{figure}[H]
  \centering
    \includegraphics[width=14cm]{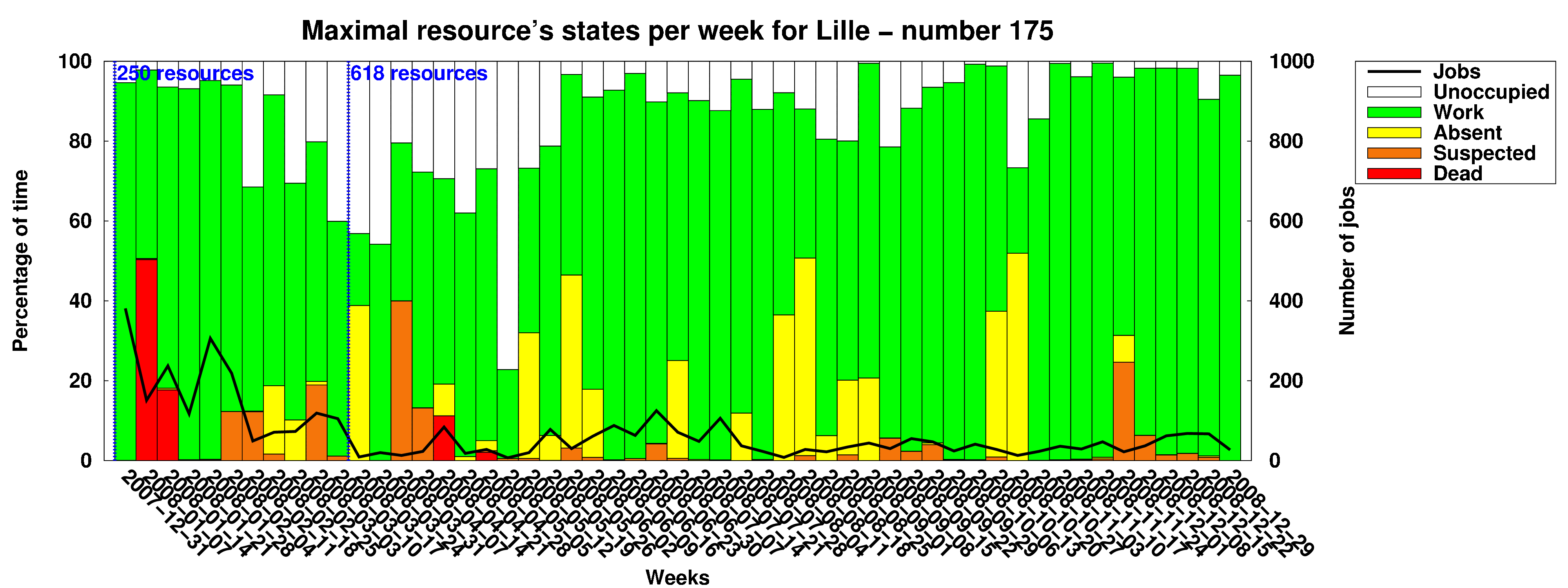}
    \caption{Maximal resource diagram for Grid'5000's Lille site}
    \label{lilmax}
\end{figure}

$\textdbend$ The number of resources has greatly increase from 250 to 618 cores during the year 2008. The last two months of the year show a really high utilization of this site (over 95\% during those two months). Moreover, we can see on Figure~\ref{lilmed} that even the oldest resources are reserved and used for experiments on Grid'5000 Lille site.

\subsection{Usage of Grid'5000's Lyon site in 2008}
\begin{itemize}
\item Platform and resources:
\begin{itemize}

\item Maximal number of resources (cores): 322
\item Mean time spent in each state for all the resources, in percentage:
\begin{itemize}
\item Dead: 1.82\%
\item Suspected: 2.78\%
\item Absent: 3.43\%
\item Work: 65.64\%
\end{itemize}
\item Real percentage of work time (without taking into account the time when the resources are dead or absent): 69.27\%
\end{itemize}

\item Jobs:
\begin{itemize}
\item Number of jobs (reservations): 138217
\item Mean time of a job: 3723.55 s. (1 hour 2 minutes and 4 seconds)
\item Maximal duration:  1080001 s. (12 days 12 hours and 1 second) for job number 124333
\item Mean number of resources (cores) per job: 4.39
\item Percentage of deploy jobs: 3.88\%
\item Percentage of time spent in deploy jobs compared to the work time: 49.69\%
\item Percentage of jobs coming from other sites: 87.07\%
\item Percentage of besteffort jobs: 44.00\%
\item Percentage of time spent in besteffort jobs compared to the work time: 24.14\%
\end{itemize}

\item Users:
\begin{itemize}
\item Number of users: 169
\item Percentage of users coming from other sites: 73.37\%
\end{itemize}
\end{itemize}

\begin{figure}[H]
  \centering
    \includegraphics[width=14cm]{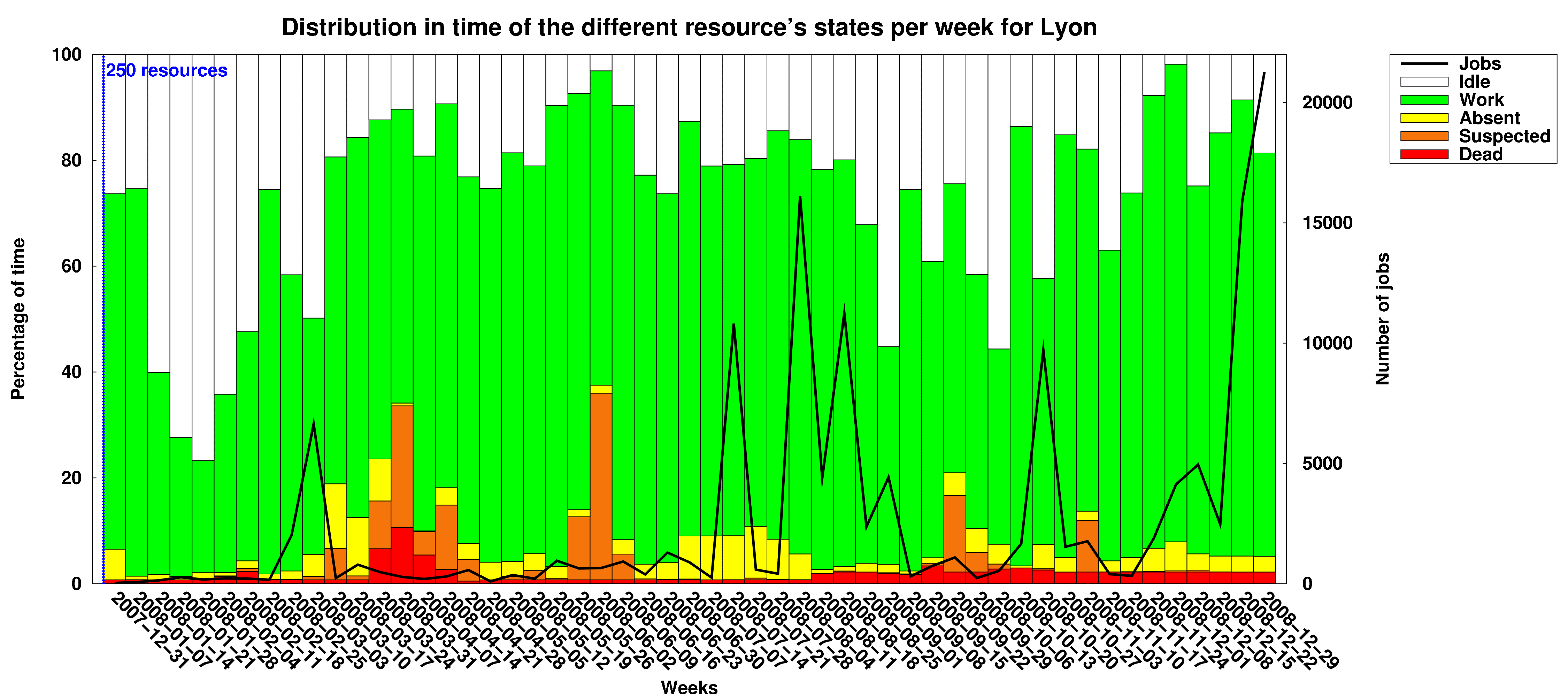}
    \caption{Global diagram with dead time for Grid'5000's Lyon site}
    \label{lyo1}
\end{figure}

\begin{figure}[H]
  \centering
   \includegraphics[width=14cm]{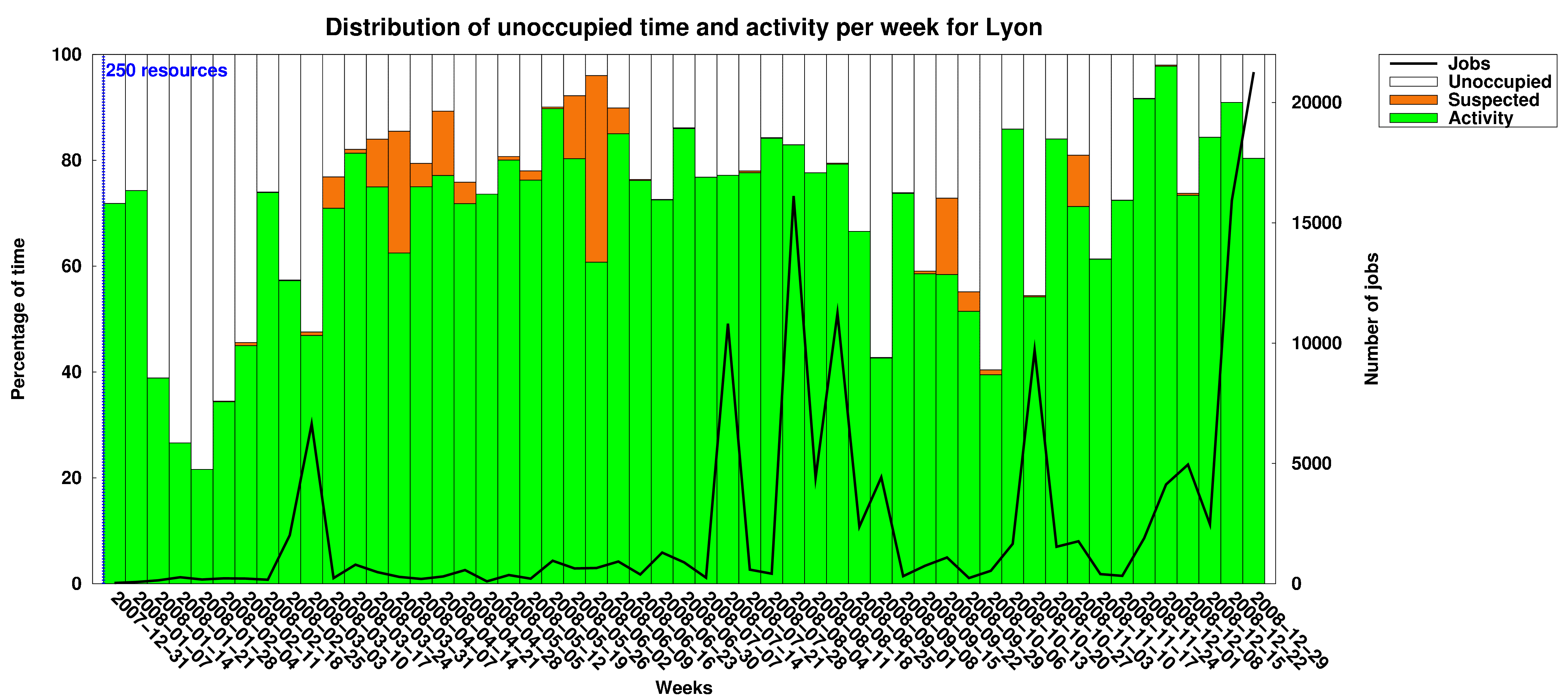}
   \caption{Global diagram without dead time for Grid'5000's Lyon site}
    \label{lyo2}
\end{figure}


\begin{figure}[H]
  \centering
    \includegraphics[width=14cm]{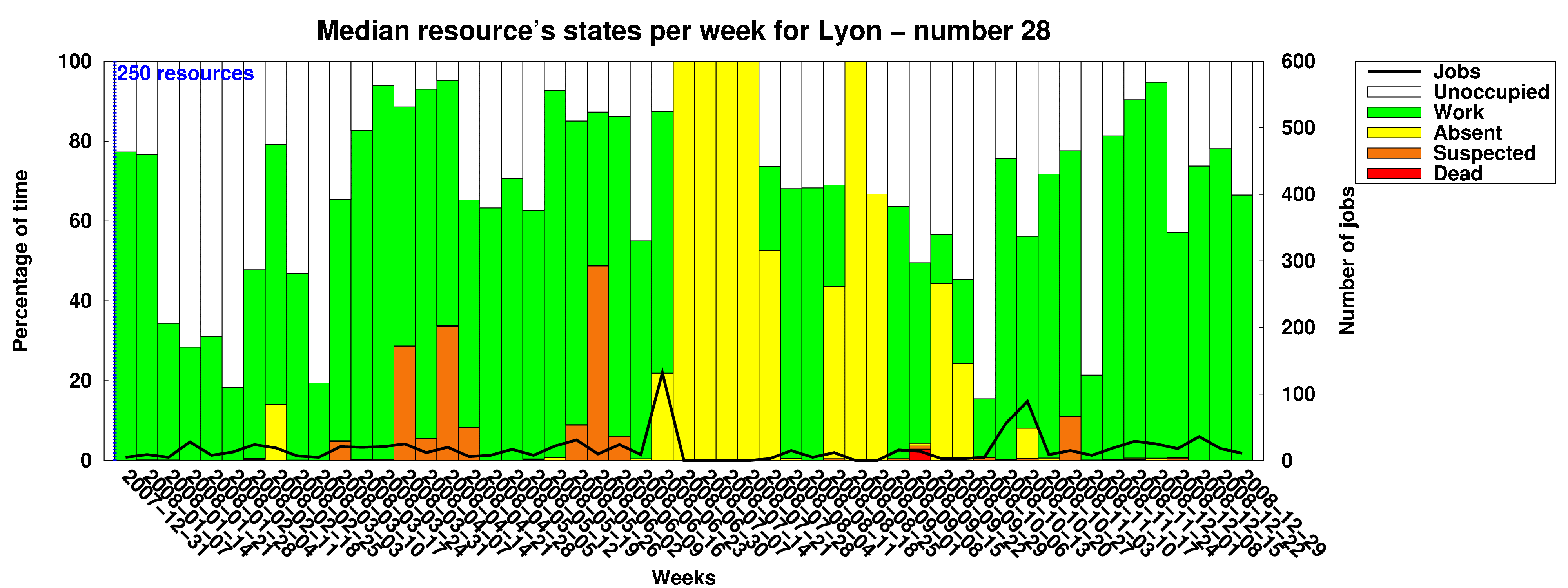}
    \caption{Median resource diagram for Grid'5000's Lyon site}
    \label{lyomed}
\end{figure}

\begin{figure}[H]
  \centering
    \includegraphics[width=14cm]{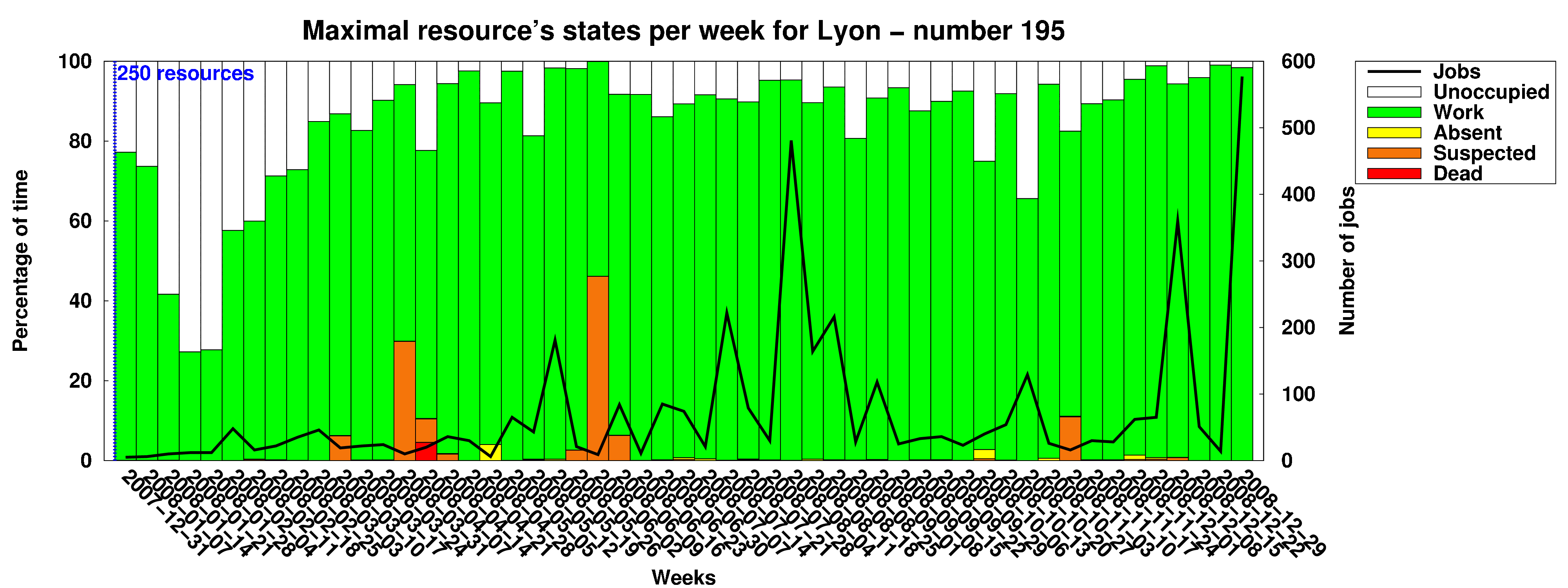}
    \caption{Maximal resource diagram for Grid'5000's Lyon site}
    \label{lyomax}
\end{figure}

$\textdbend$ Lyon is the smallest Grid'5000 site for the year 2008. No machine has been added during this year to this site. However, it is really used. Another peculiarity of this site is that it gets the smallest reservations in terms of mean number of resources per reservation and among the smallest reservations in terms of mean length. It also gets the biggest time spent (in percentage) in deploy jobs: almost half of the time nodes are working with user's environments.

\subsection{Usage of Grid'5000's Nancy site in 2008}
\begin{itemize}
\item Platform and resources:
\begin{itemize}

\item Maximal number of resources (cores): 574
\item Mean time spent in each state for all the resources, in percentage:
\begin{itemize}
\item Dead: 0.11\%
\item Suspected: 0.90\%
\item Absent: 1.95\%
\item Work: 58.84\%
\end{itemize}
\item Real percentage of work time (without taking into account the time when the resources are dead or absent): 60.08\%
\end{itemize}

\item Jobs:
\begin{itemize}
\item Number of jobs (reservations): 74592
\item Mean time of a job: 8912.82 s. (2 hours 28 minutes and 33 seconds)
\item Maximal duration:  1695503 s. (19 days 14 hours 58 minutes and 23 seconds) for job number 213465
\item Mean number of resources (cores) per job: 14.63
\item Percentage of deploy jobs: 4.23\%
\item Percentage of time spent in deploy jobs compared to the work time: 29.95\%
\item Percentage of jobs coming from other sites: 20.11\%
\item Percentage of besteffort jobs: 76.16\%
\item Percentage of time spent in besteffort jobs compared to the work time: 17.05\%
\end{itemize}

\item Users:
\begin{itemize}
\item Number of users: 157
\item Percentage of users coming from other sites: 83.44\%
\end{itemize}
\end{itemize}

\begin{figure}[H]
  \centering
    \includegraphics[width=14cm]{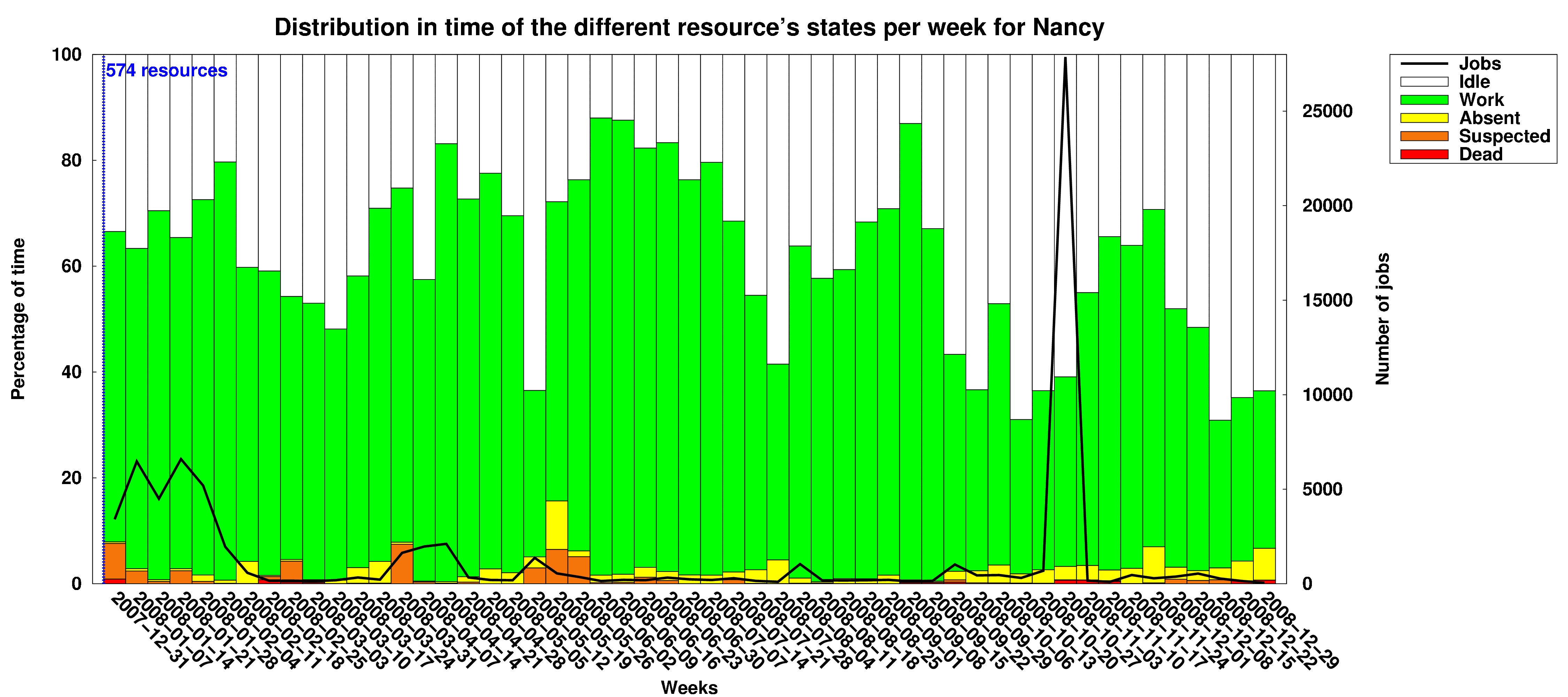}
    \caption{Global diagram with dead time for Grid'5000's Nancy site}
    \label{nan1}
\end{figure}

\begin{figure}[H]
  \centering
   \includegraphics[width=14cm]{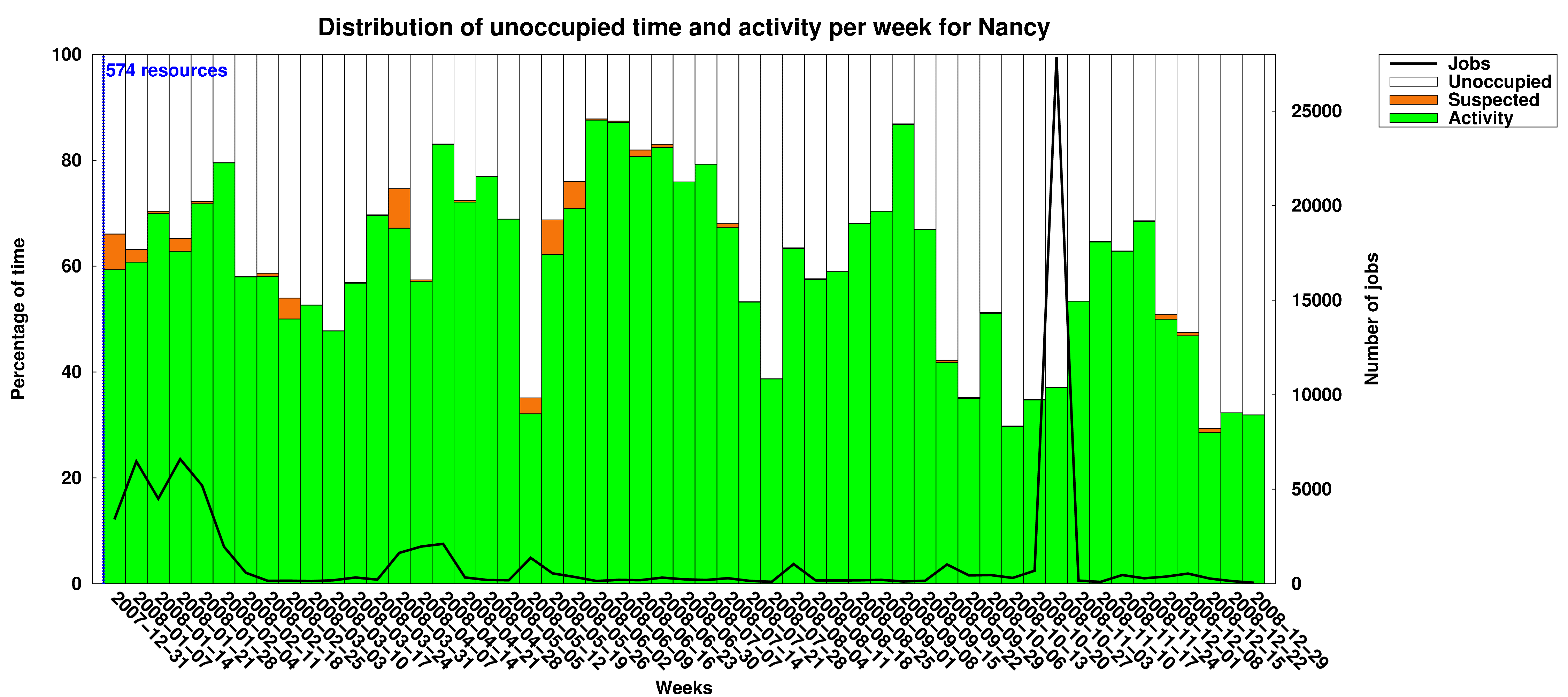}
   \caption{Global diagram without dead time for Grid'5000's Nancy site}
    \label{nan2}
\end{figure}


\begin{figure}[H]
  \centering
    \includegraphics[width=14cm]{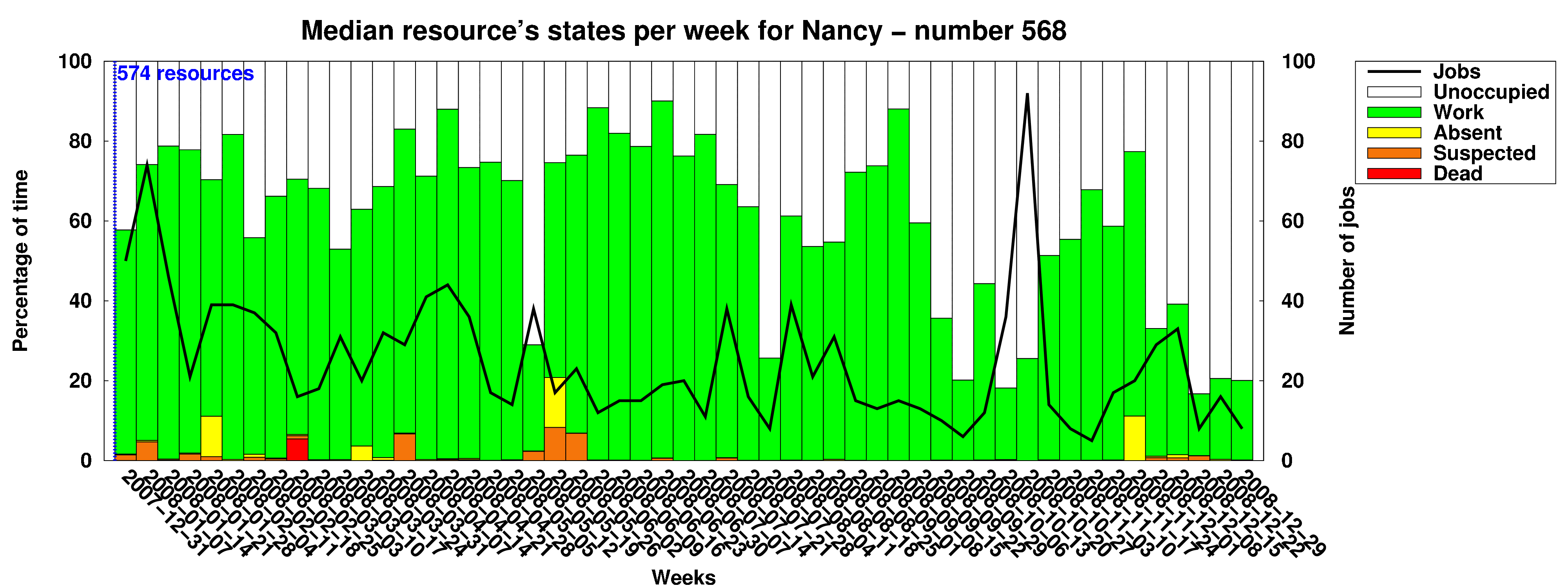}
    \caption{Median resource diagram for Grid'5000's Nancy site}
    \label{nanmed}
\end{figure}

\begin{figure}[H]
  \centering
    \includegraphics[width=14cm]{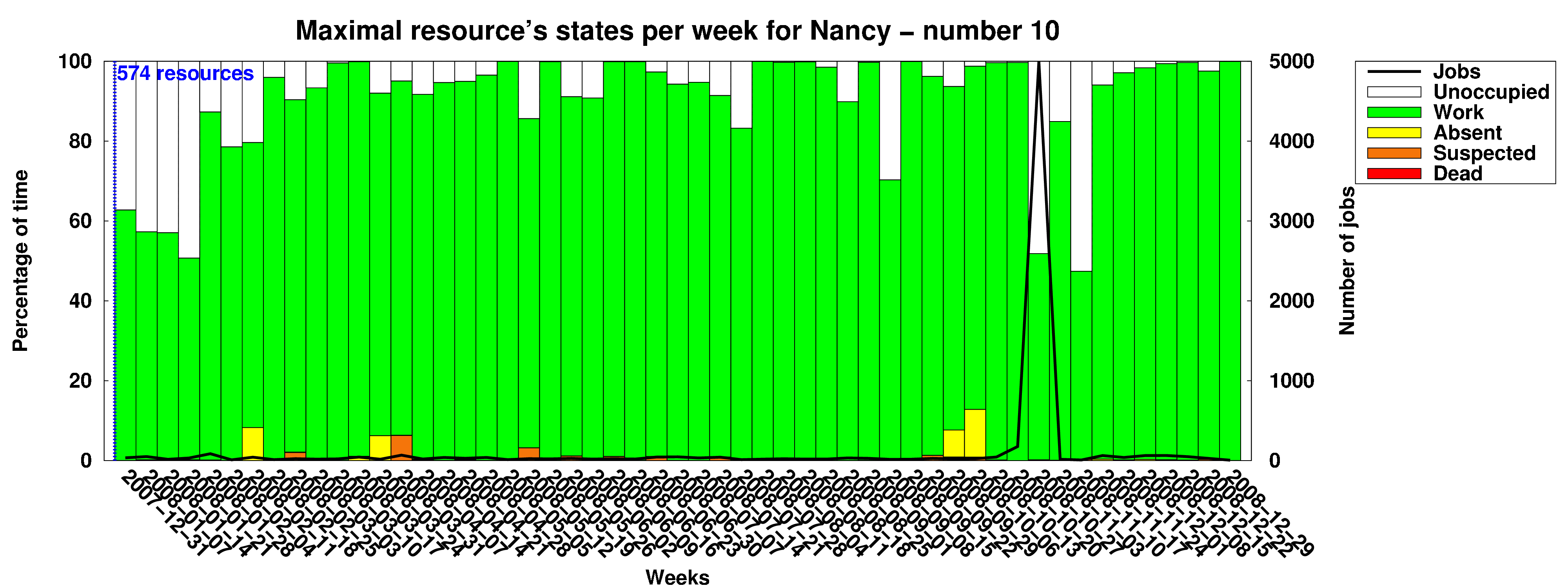}
    \caption{Maximal resource diagram for Grid'5000's Nancy site}
    \label{nanmax}
\end{figure}

$\textdbend$ The amount of available resources remains constant during the year 2008. This site presents a healthy working: almost no dead, suspected and absent periods. However, the load is not really balanced between the nodes as shown by Figures~\ref{nanmed} and~\ref{nanmax}.

\subsection{Usage of Grid'5000's Orsay site in 2008}
\begin{itemize}
\item Platform and resources:
\begin{itemize}

\item Maximal number of resources (cores): 684
\item Mean time spent in each state for all the resources, in percentage:
\begin{itemize}
\item Dead: 5.82\%
\item Suspected: 12.20\%
\item Absent: 6.45\%
\item Work: 50.73\%
\end{itemize}
\item Real percentage of work time (without taking into account the time when the resources are dead or absent): 57.82\%
\end{itemize}

\item Jobs:
\begin{itemize}
\item Number of jobs (reservations): 92862
\item Mean time of a job: 6246.07 s. (1 hour 44 minutes and 6 seconds)
\item Maximal duration:  1060487 s. (12 days 6 hours 34 minutes and 47 seconds) for job number 101189
\item Mean number of resources (cores) per job: 14.58
\item Percentage of deploy jobs: 3.59\%
\item Percentage of time spent in deploy jobs compared to the work time: 38.98\%
\item Percentage of jobs coming from other sites: 96.28\%
\item Percentage of besteffort jobs: 38.91\%
\item Percentage of time spent in besteffort jobs compared to the work time: 27.17\%
\end{itemize}

\item Users:
\begin{itemize}
\item Number of users: 172
\item Percentage of users coming from other sites: 76.16\%
\end{itemize}
\end{itemize}

\begin{figure}[H]
  \centering
    \includegraphics[width=14cm]{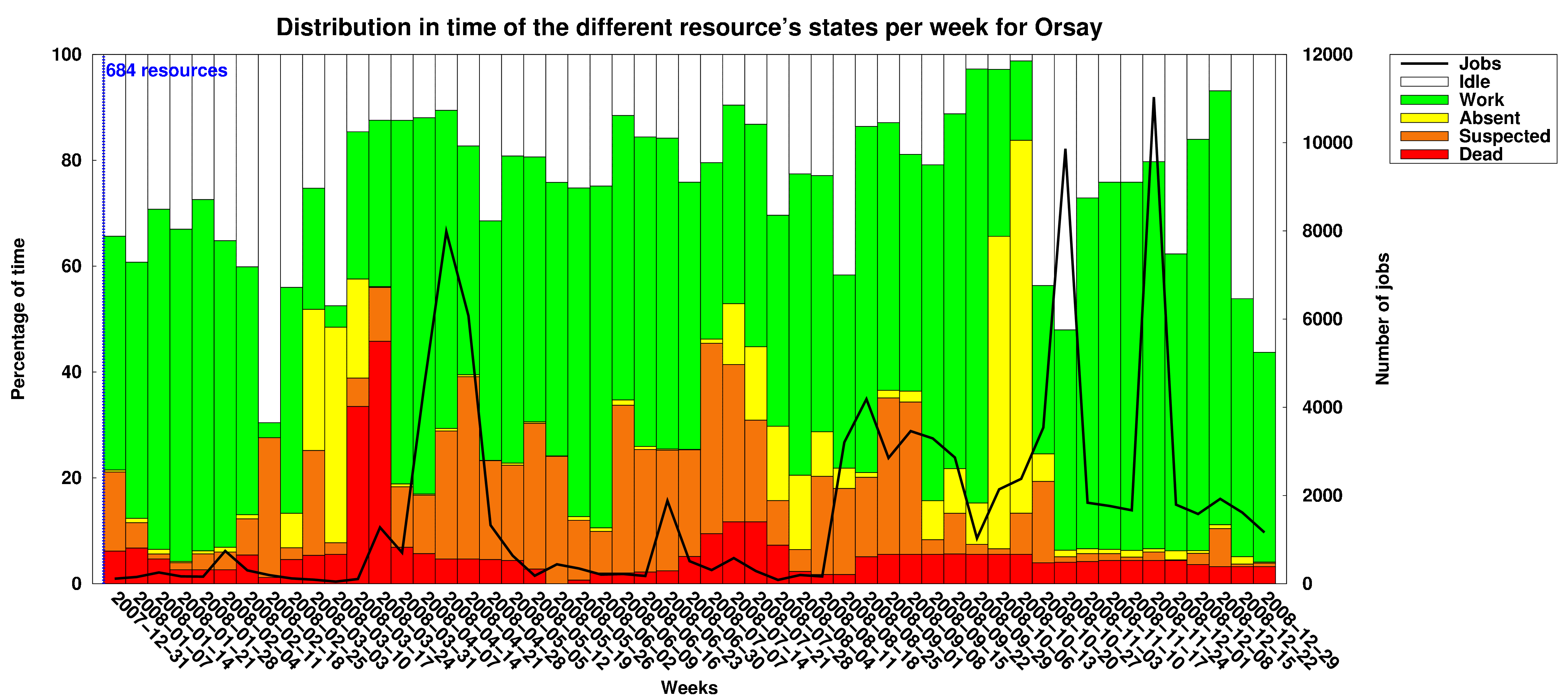}
    \caption{Global diagram with dead time for Grid'5000's Orsay site}
    \label{ors1}
\end{figure}

\begin{figure}[H]
  \centering
   \includegraphics[width=14cm]{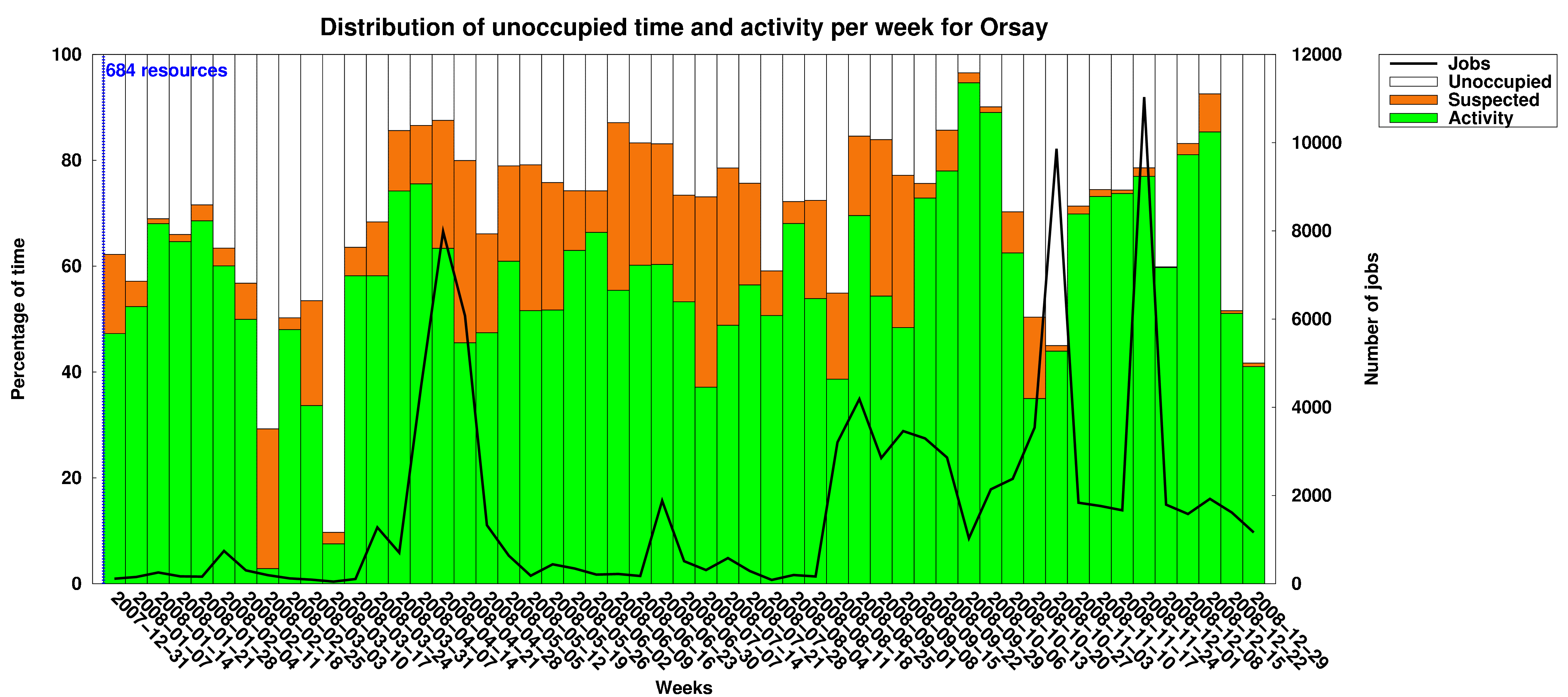}
   \caption{Global diagram without dead time for Grid'5000's Orsay site}
    \label{ors2}
  
\end{figure}


\begin{figure}[H]
  \centering
    \includegraphics[width=14cm]{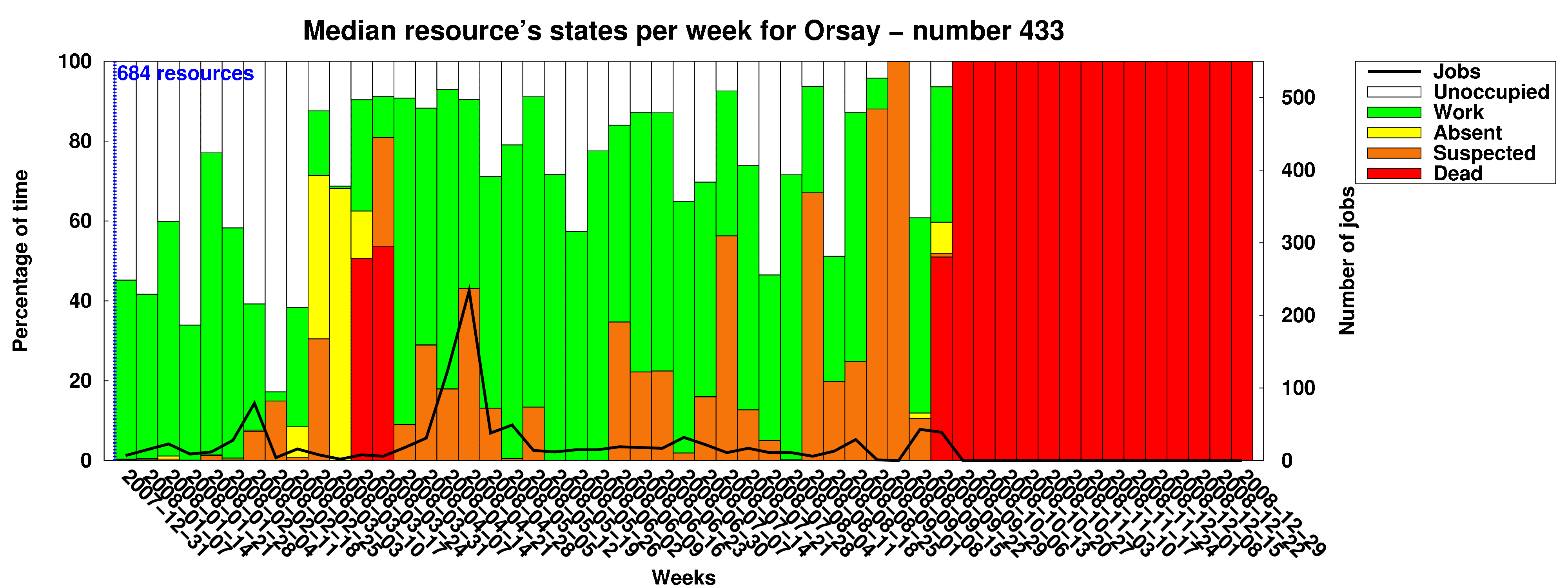}
    \caption{Median resource diagram for Grid'5000's Orsay site}
    \label{orsmed}
\end{figure}

\begin{figure}[H]
  \centering
    \includegraphics[width=14cm]{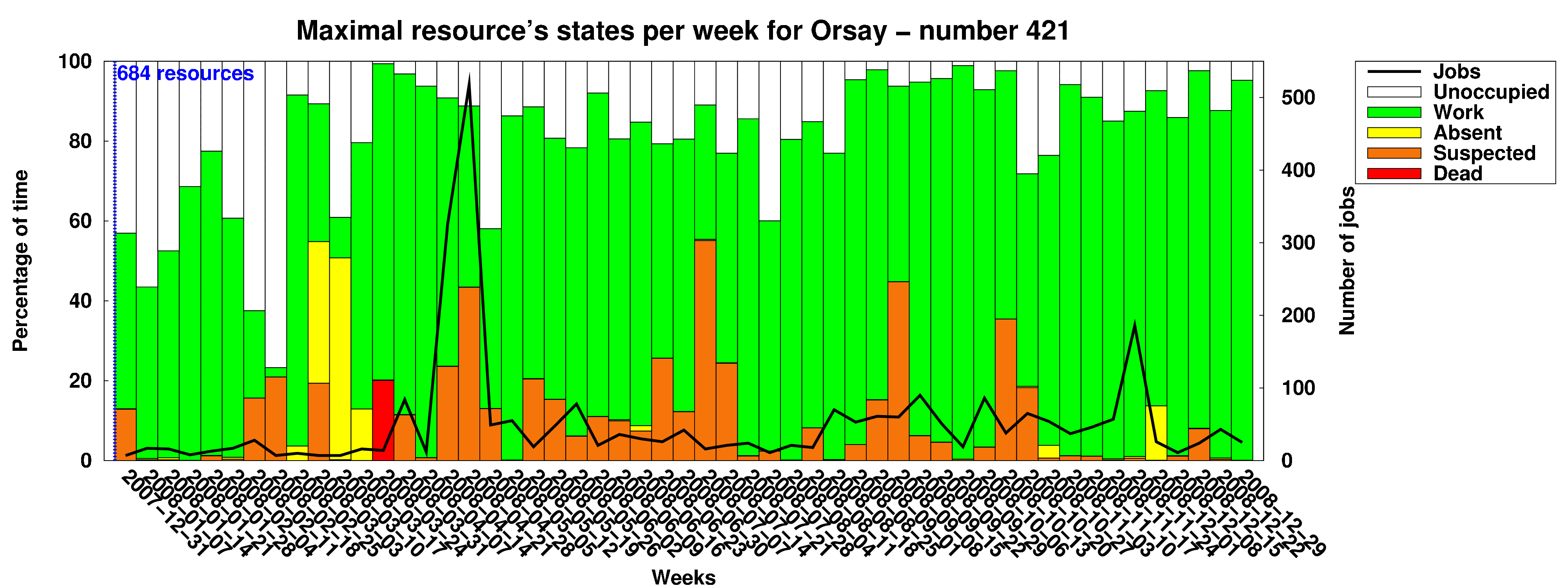}
    \caption{Maximal resource diagram for Grid'5000's Orsay site}
    \label{orsmax}
\end{figure}

$\textdbend$ This site is really more used than the previous year (18.88\% of real activity in 2007~\cite{rapport2007}). It is one of the biggest sites in terms of resources and attracts a lot of jobs coming from other sites.

\subsection{Usage of Grid'5000's Rennes site in 2008}
\begin{itemize}
\item Platform and resources:
\begin{itemize}

\item Maximal number of resources (cores): 714
\item Mean time spent in each state for all the resources, in percentage:
\begin{itemize}
\item Dead: 4.79\%
\item Suspected: 0.44\%
\item Absent: 0.79\%
\item Work: 60.98\%
\end{itemize}
\item Real percentage of work time (without taking into account the time when the resources are dead or absent): 64.58\%
\end{itemize}

\item Jobs:
\begin{itemize}
\item Number of jobs (reservations): 58843
\item Mean time of a job: 7069.33 s. (1 hour 57 minutes and 49 seconds)
\item Maximal duration:  604825 s. (7 days and 25 seconds) for job number 289263
\item Mean number of resources (cores) per job: 27.32
\item Percentage of deploy jobs: 7.84\%
\item Percentage of time spent in deploy jobs compared to the work time: 35.96\%
\item Percentage of jobs coming from other sites: 87.41\%
\item Percentage of besteffort jobs: 50.01\%
\item Percentage of time spent in besteffort jobs compared to the work time: 23.38\%
\end{itemize}

\item Users:
\begin{itemize}
\item Number of users: 213
\item Percentage of users coming from other sites: 82.63\%
\end{itemize}
\end{itemize}

\begin{figure}[H]
  \centering
    \includegraphics[width=14cm]{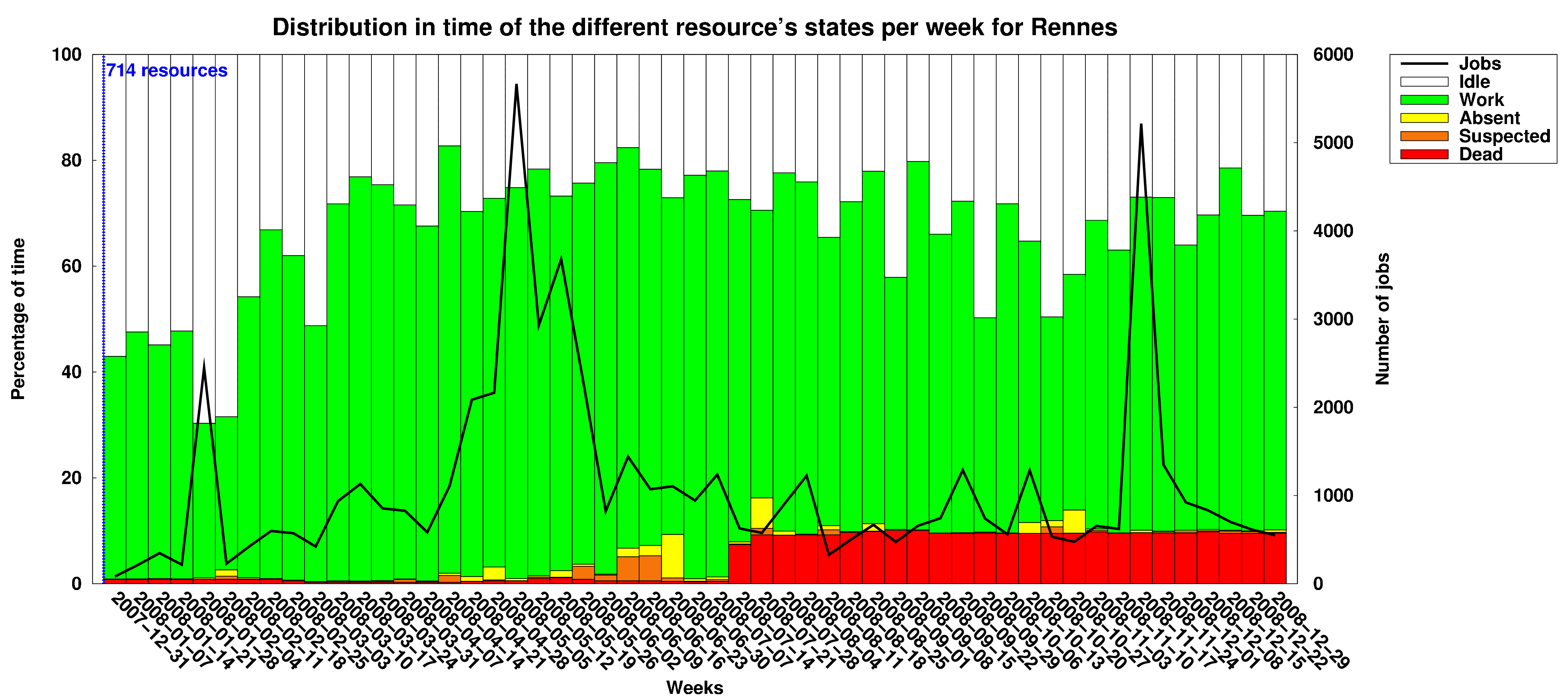}
    \caption{Global diagram with dead time for Grid'5000's Rennes site}
    \label{ren1}
\end{figure}

\begin{figure}[H]
  \centering
   \includegraphics[width=14cm]{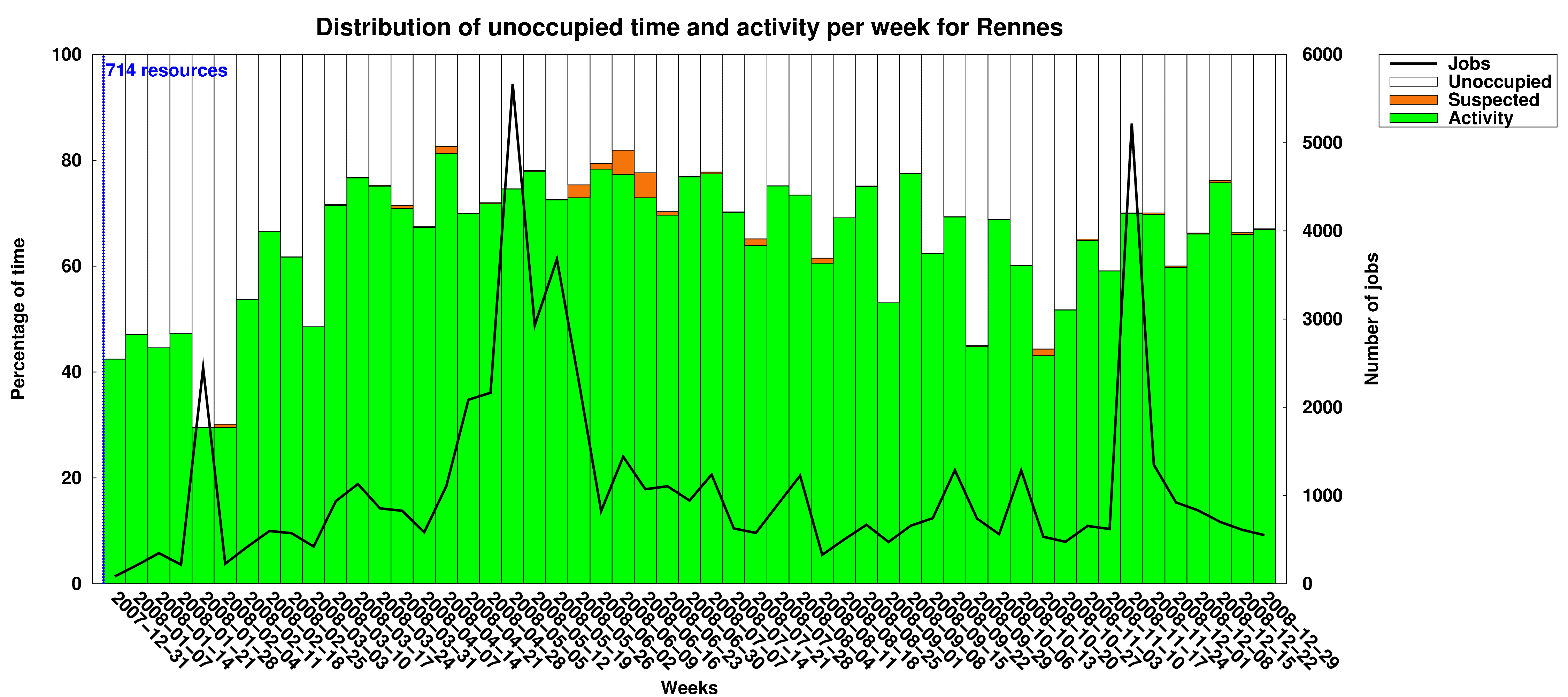}
   \caption{Global diagram without dead time for Grid'5000's Rennes site}
    \label{ren2}
\end{figure}


\begin{figure}[H]
  \centering
    \includegraphics[width=14cm]{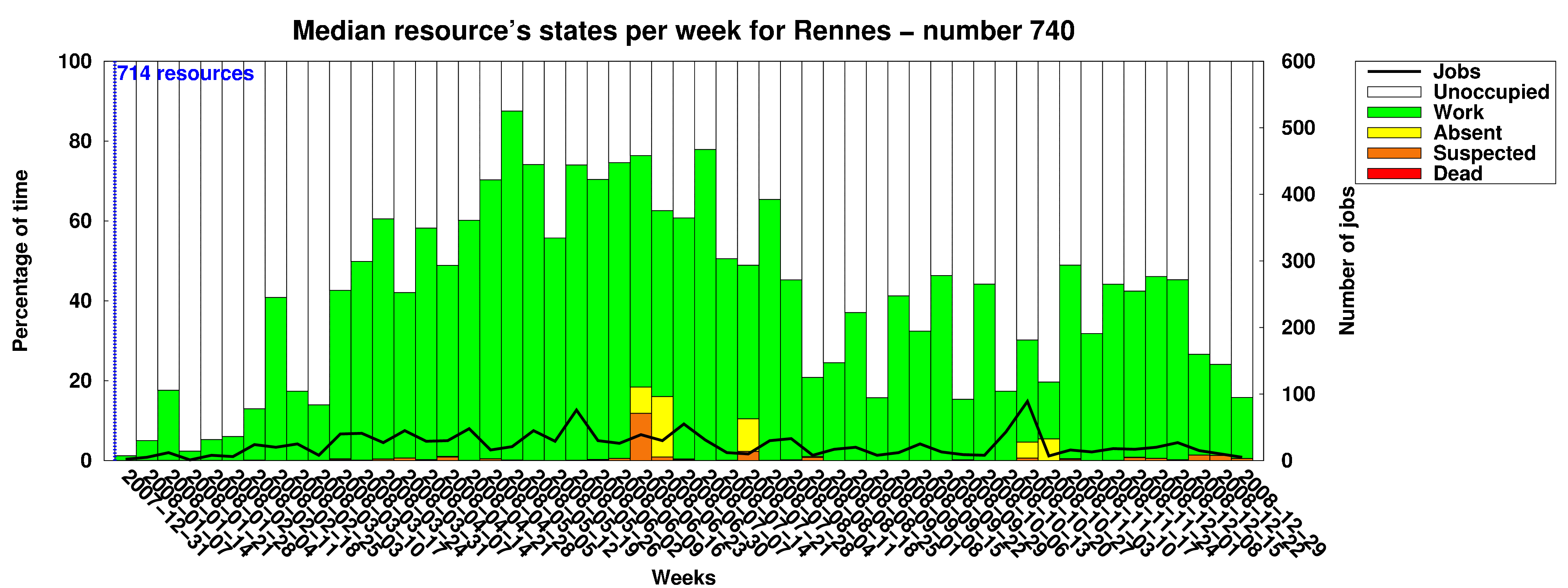}
    \caption{Median resource diagram for Grid'5000's Rennes site}
    \label{renmed}
\end{figure}

\begin{figure}[H]
  \centering
    \includegraphics[width=14cm]{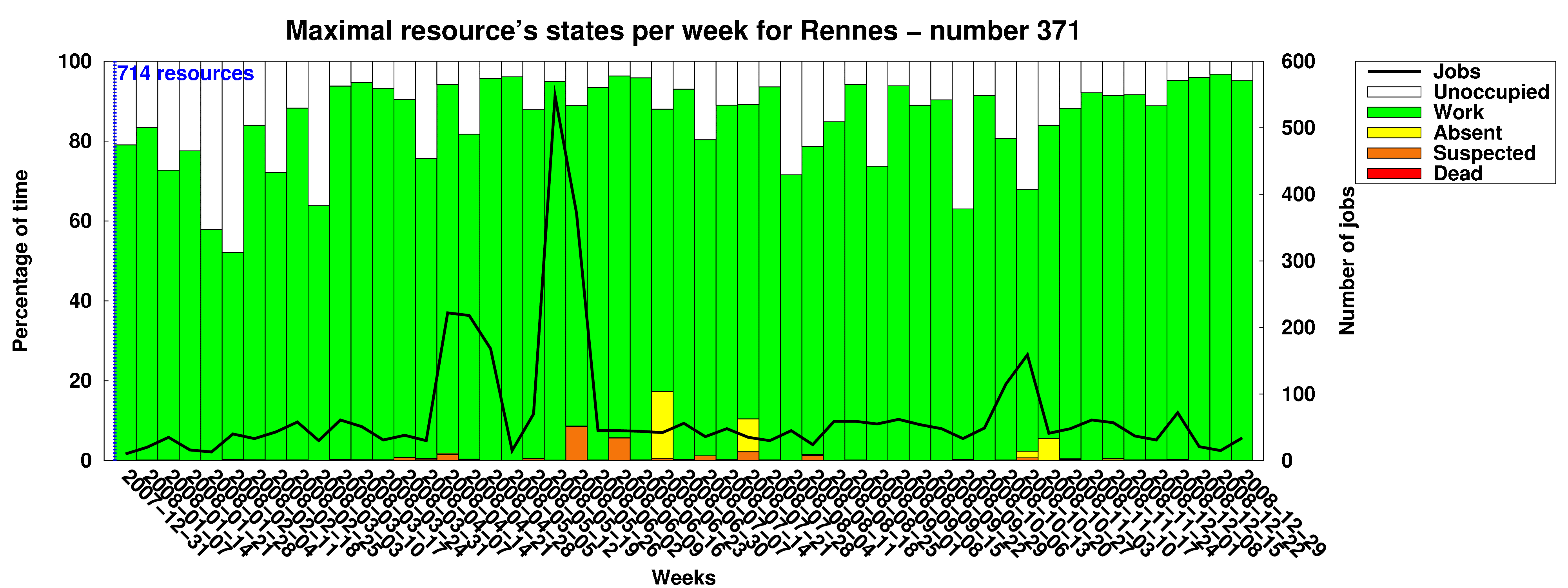}
    \caption{Maximal resource diagram for Grid'5000's Rennes site}
    \label{renmax}
\end{figure}

$\textdbend$ From Figures~\ref{renmed} and~\ref{renmax}, we see a really heterogeneous usage of the resources (probably) due to the heterogeneity of the nodes. This site is the biggest one in terms of resources and its activity percentage is really on the average. Due to its size, it allows bigger reservation. So, it has the higher mean number of resources per reservation (27.32) but one of the smallest number of reservations for the whole year.

\subsection{Usage of Grid'5000's Sophia site in 2008}
\begin{itemize}
\item Platform and resources:
\begin{itemize}

\item Maximal number of resources (cores): 568
\item Mean time spent in each state for all the resources, in percentage:
\begin{itemize}
\item Dead: 2.04\%
\item Suspected: 0.37\%
\item Absent: 1.98\%
\item Work: 78.25\%
\end{itemize}
\item Real percentage of work time (without taking into account the time when the resources are dead or absent): 81.51\%
\end{itemize}

\item Jobs:
\begin{itemize}
\item Number of jobs (reservations): 58142
\item Mean time of a job: 8767.35 s. (2 hours 26 minutes and 7 seconds)
\item Maximal duration:  485924 s. (5 days 14 hours 58 minutes and 44 seconds) for job number 345959
\item Mean number of resources (cores) per job: 22.14
\item Percentage of deploy jobs: 8.76\%
\item Percentage of time spent in deploy jobs compared to the work time: 25.64\%
\item Percentage of jobs coming from other sites: 87.24\%
\item Percentage of besteffort jobs: 54.36\%
\item Percentage of time spent in besteffort jobs compared to the work time: 19.57\%
\end{itemize}

\item Users:
\begin{itemize}
\item Number of users: 187
\item Percentage of users coming from other sites: 85.03\%
\end{itemize}
\end{itemize}

\begin{figure}[H]
  \centering
    \includegraphics[width=14cm]{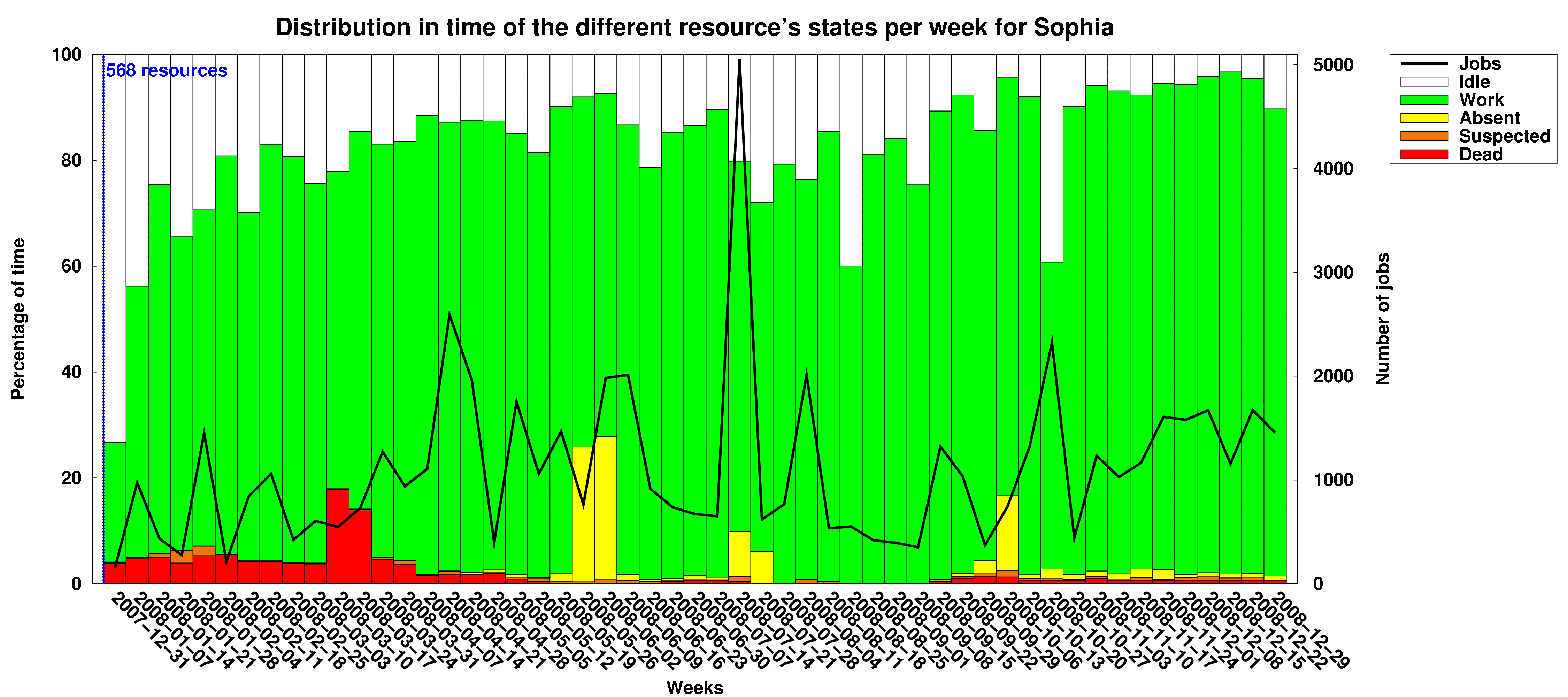}
    \caption{Global diagram with dead time for Grid'5000's Sophia site}
    \label{sop1}
\end{figure}

\begin{figure}[H]
  \centering
   \includegraphics[width=14cm]{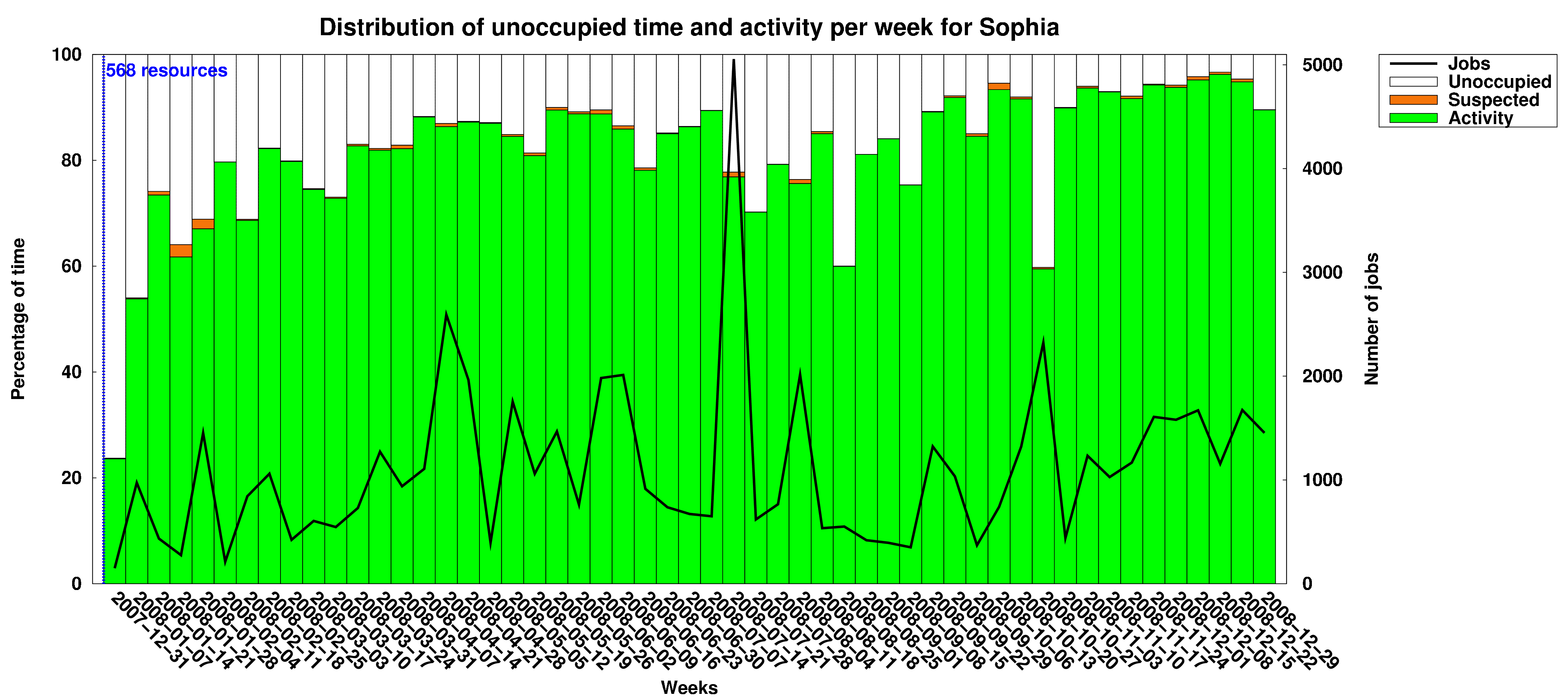}
   \caption{Global diagram without dead time for Grid'5000's Sophia site}
    \label{sop2}
\end{figure}


\begin{figure}[H]
  \centering
    \includegraphics[width=14cm]{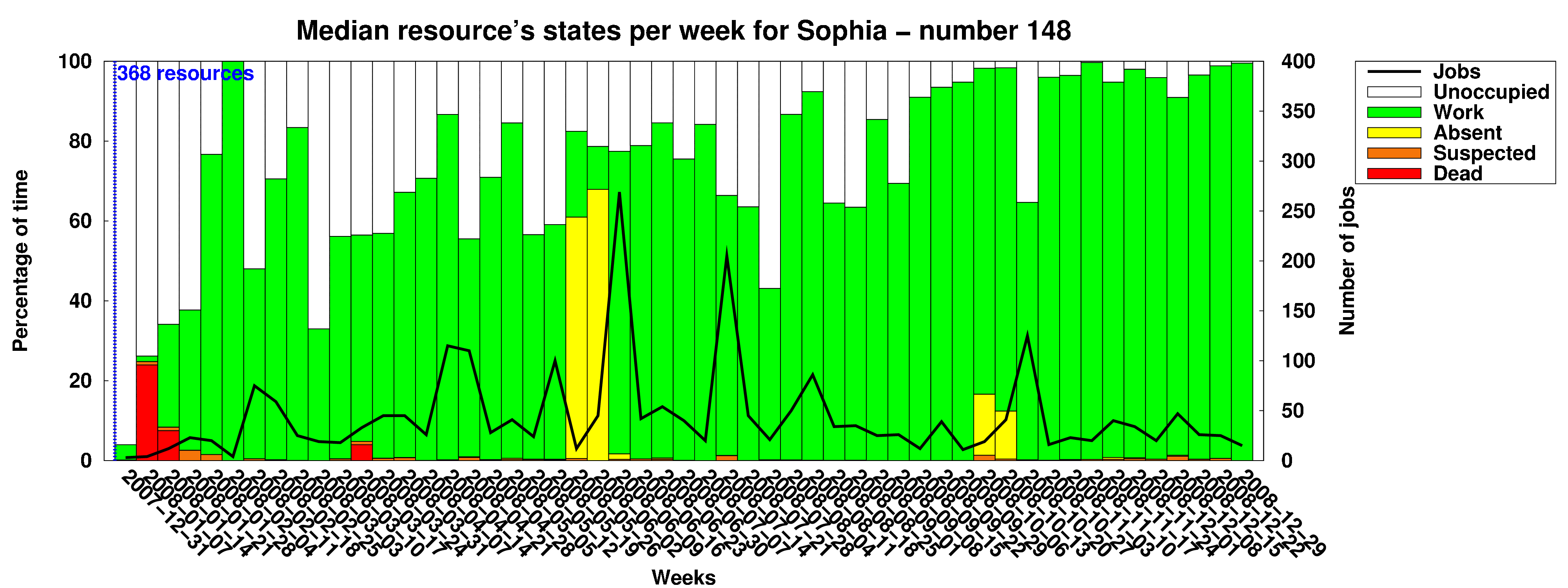}
    \caption{Median resource diagram for Grid'5000's Sophia site}
    \label{sopmed}
\end{figure}

\begin{figure}[H]
  \centering
    \includegraphics[width=14cm]{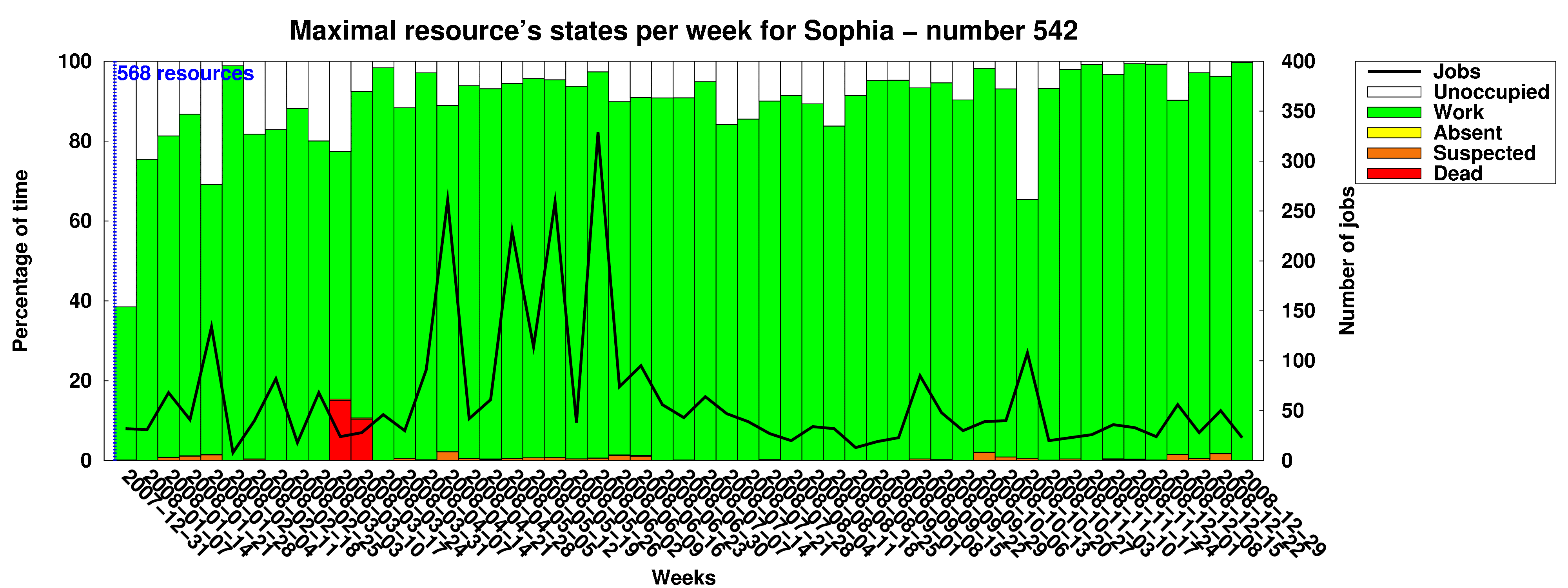}
    \caption{Maximal resource diagram for Grid'5000's Sophia site}
    \label{sopmax}
\end{figure}

$\textdbend$ This site has the smallest number of jobs for the whole year compared to the other Grid'5000 sites. But, due to a high mean number of resources per reservation, it gets the highest percentage of activity (81.51\%). This is illustrated on Figures~\ref{sop1} and~\ref{sop2}.

\subsection{Usage of Grid'5000's Toulouse site in 2008}
\begin{itemize}
\item Platform and resources:
\begin{itemize}

\item Maximal number of resources (cores): 434
\item Mean time spent in each state for all the resources, in percentage:
\begin{itemize}
\item Dead: 5.00\%
\item Suspected: 1.22\%
\item Absent: 14.86\%
\item Work: 49.42\%
\end{itemize}
\item Real percentage of work time (without taking into account the time when the resources are dead or absent): 61.67\%
\end{itemize}

\item Jobs:
\begin{itemize}
\item Number of jobs (reservations): 166191
\item Mean time of a job: 2211.80 s. (36 minutes and 52 seconds)
\item Maximal duration:  1080018 s. (12 days 12 hours and 18 seconds) for job number 94101
\item Mean number of resources (cores) per job: 6.29
\item Percentage of deploy jobs: 1.47\%
\item Percentage of time spent in deploy jobs compared to the work time: 28.37\%
\item Percentage of jobs coming from other sites: 96.15\%
\item Percentage of besteffort jobs: 6.86\%
\item Percentage of time spent in besteffort jobs compared to the work time: 24.85\%
\end{itemize}

\item Users:
\begin{itemize}
\item Number of users: 160
\item Percentage of users coming from other sites: 81.88\%
\end{itemize}
\end{itemize}

\begin{figure}[H]
  \centering
    \includegraphics[width=14cm]{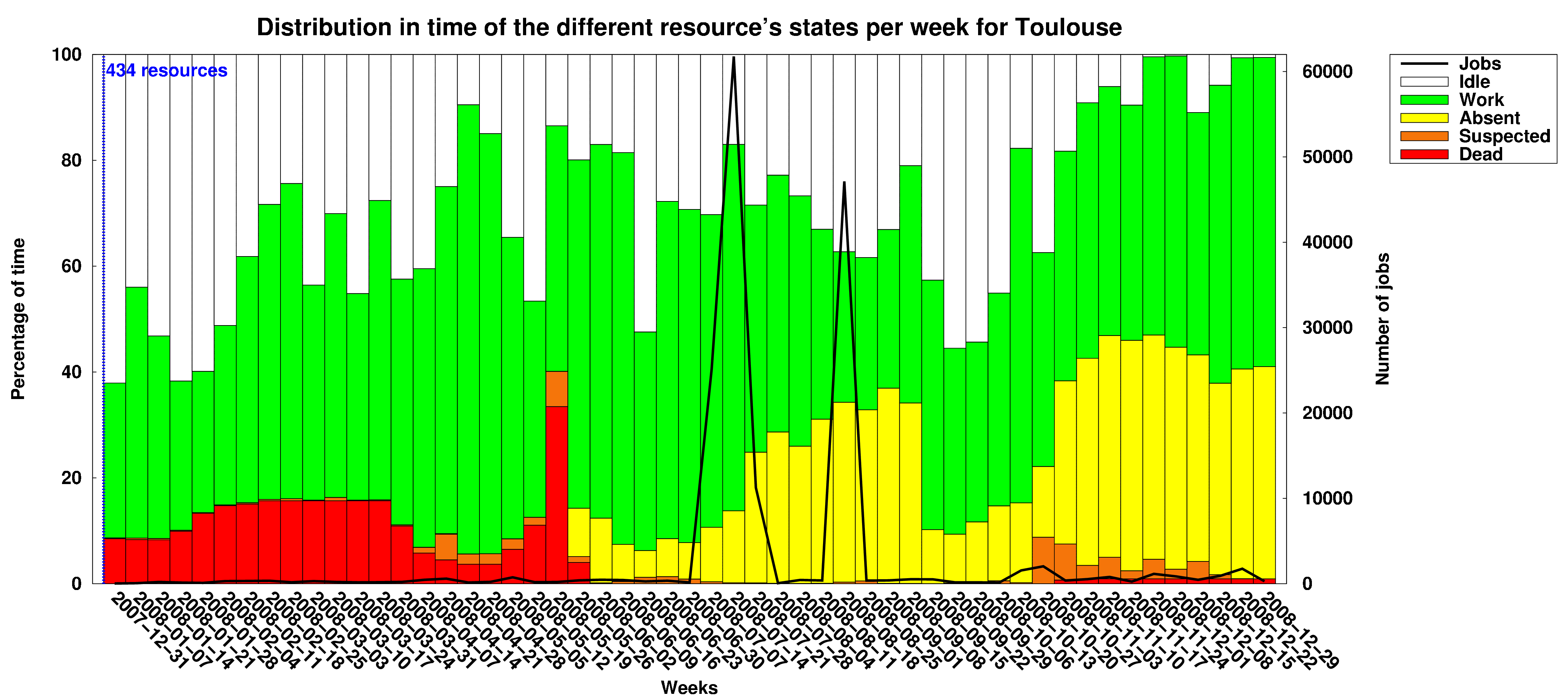}
    \caption{Global diagram with dead time for Grid'5000's Toulouse site}
    \label{tlse1}
\end{figure}

\begin{figure}[H]
  \centering
   \includegraphics[width=14cm]{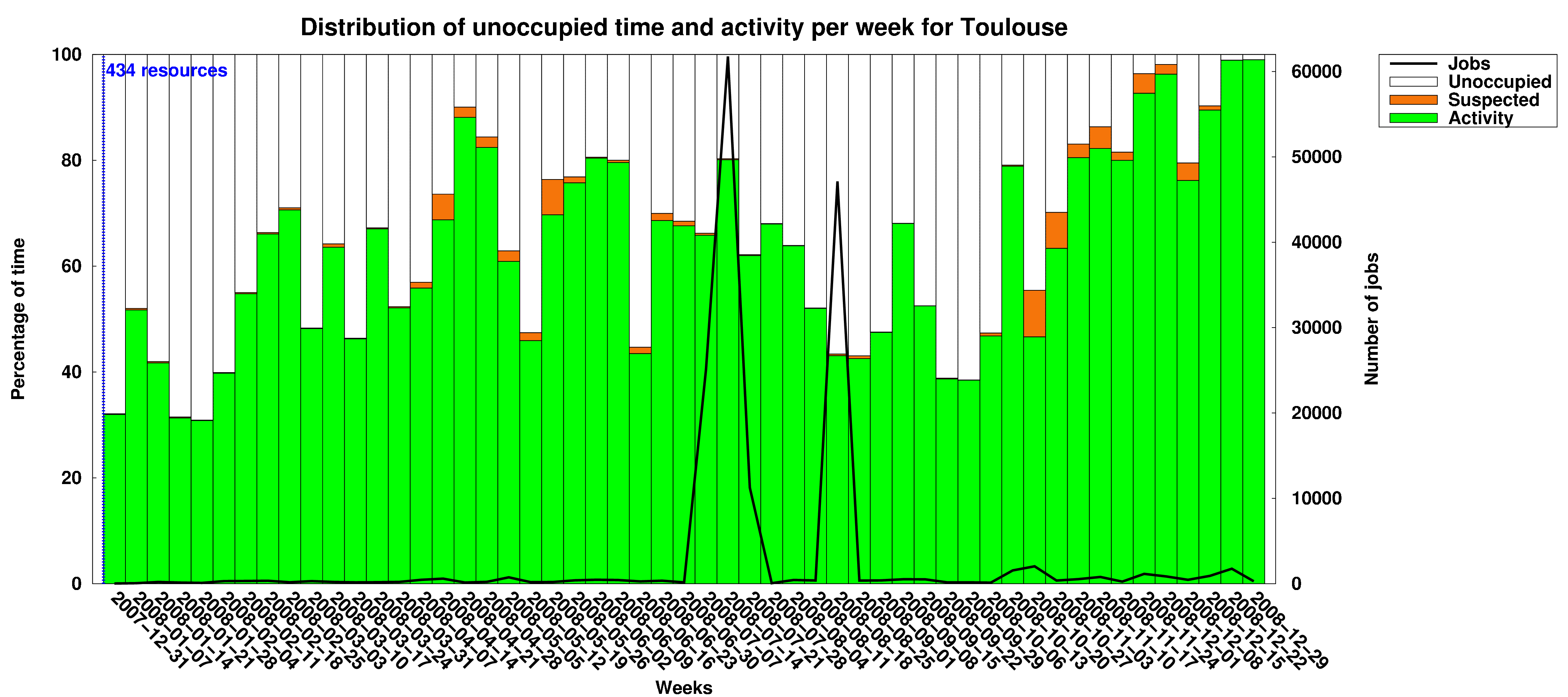}
   \caption{Global diagram without dead time for Grid'5000's Toulouse site}
    \label{tlse2}
\end{figure}


\begin{figure}[H]
  \centering
    \includegraphics[width=14cm]{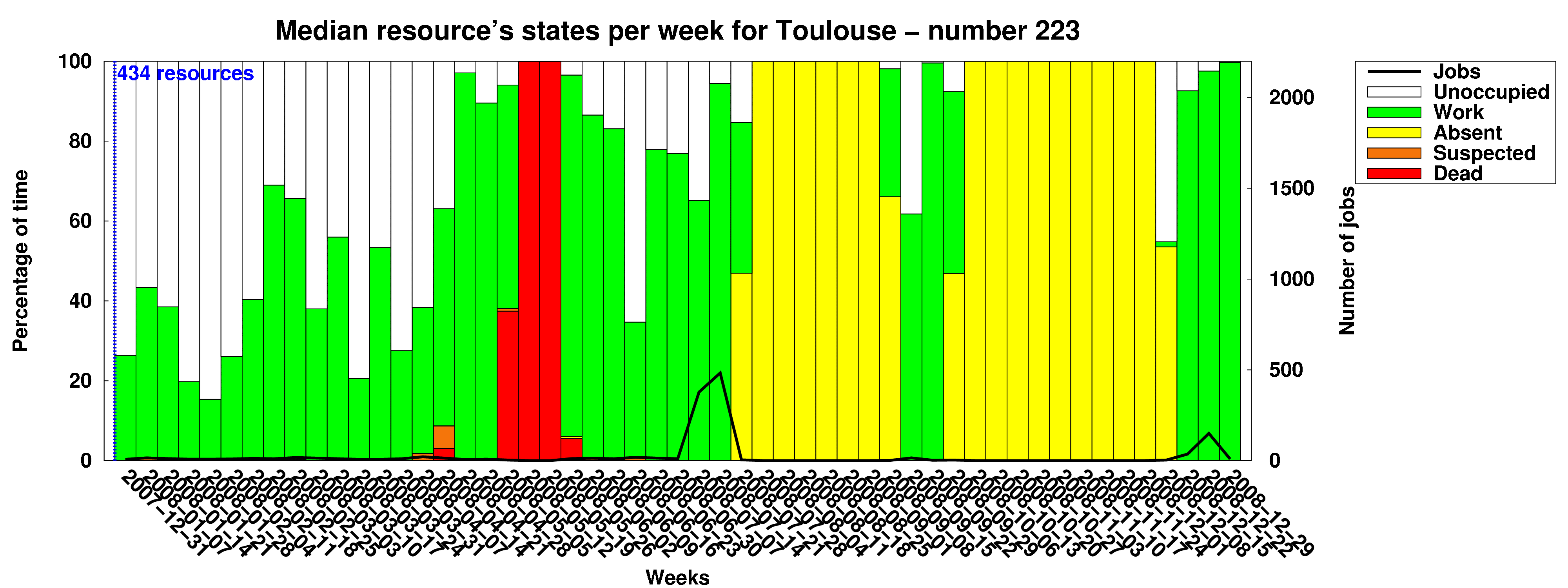}
    \caption{Median resource diagram for Grid'5000's Toulouse site}
    \label{tlsemed}
\end{figure}

\begin{figure}[H]
  \centering
    \includegraphics[width=14cm]{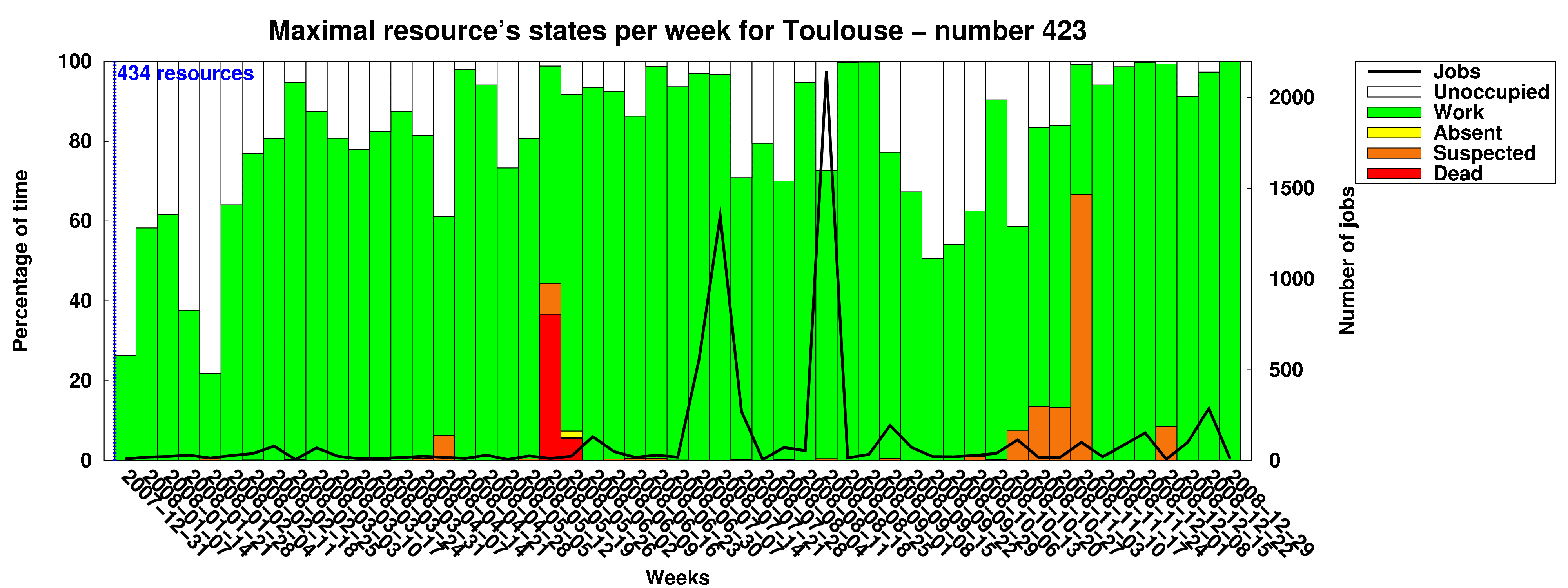}
    \caption{Maximal resource diagram for Grid'5000's Toulouse site}
    \label{tlsemax} 
\end{figure}

$\textdbend$ Toulouse has the shortest reservations with an average around 36 minutes. During the last two weeks of the year, the site is used at its full capacity (about 98\% of activity). This site is one of the smallest Grid'5000 sites.

\subsection{Results summary}

Tables~\ref{res_stat},~\ref{jobs_stat},~\ref{jobs_stat2},~\ref{users_stat}, summarize the main  observed statistics in terms of resources, jobs and users. These results show that the Grid5000 platform faces some heterogeneous usage depending on the involved sites.

\begin{itemize}
\item Platform and resources:

\begin{table}[H]
\centering
\begin{tabular}{|c||c|c|}
\hline
Site & Number of resources (cores)&  Percentage of `real' activity\\ \hline \hline
Bordeaux &  650 &  53.20\% \\ \hline
Lille    &  618 &  72.89\% \\ \hline
Lyon     &  322 &  69.27\% \\ \hline
Nancy    &  574 &  60.08\% \\ \hline
Orsay    &  684 &  57.82\% \\ \hline
Rennes   &  714 &  64.58\% \\ \hline
Sophia   &  568 &  81.51\% \\ \hline
Toulouse &  434 &  61.67\% \\ \hline \hline
Total Grid'5000 & 4564 & 64.67\% \\ \hline
\end{tabular}
\caption{Resource-related statistics}
\label{res_stat}
\end{table}

\item Jobs:

\begin{table}[H]
\centering
\renewcommand{\multirowsetup}{\centering}
\begin{tabular}{|c||c|c|c|c|}
\hline
\multirow{3}*{Site}
& Number of & Mean number& Mean duration& Jobs coming \\
& jobs & of resources& of a job & from other \\
& (reservations) & per job & (seconds) & sites \\
\hline \hline
Bordeaux &  356222 &   7.44 &  2473.38 &  83.28\% \\ \hline
Lille    &  344538 &   8.11 &  3154.58 &  38.97\% \\ \hline
Lyon     &  138217 &   4.39 &  3723.55 &  87.07\% \\ \hline
Nancy    &   74592 &  14.63 &  8912.82 &  20.11\% \\ \hline
Orsay    &   92862 &  14.58 &  6246.07 &  96.28\% \\ \hline
Rennes   &   58843 &  27.32 &  7069.33 &  87.41\% \\ \hline
Sophia   &   58142 &  22.14 &  8767.35 &  87.24\% \\ \hline
Toulouse &  166191 &   6.29 &  2211.80 &  96.15\% \\ \hline
\end{tabular}
\caption{Job-related statistics}
\label{jobs_stat}
\end{table}

\begin{table}[H]
\centering
\renewcommand{\multirowsetup}{\centering}
\begin{tabular}{|c||c|c|c|c|c|c|}
\hline
\multirow{3}*{Site}
& Percentage & Time spent & Percentage & Time spent & Percentage & Time spent\\
& of deploy & in deploy & of besteffort & in besteffort & of default & in default\\
& jobs & jobs & jobs & jobs & jobs & jobs\\
\hline \hline
Bordeaux &  0.97\% & 22.33\% & 58.03\% & 34.90\% & 41.00\% & 42.77\% \\ \hline
Lille    &  0.61\% &  7.24\% & 19.10\% & 37.88\% & 80.29\% & 54.88\% \\ \hline
Lyon     &  3.88\% & 49.69\% & 44.00\% & 24.14\% & 52.12\% & 26.17\% \\ \hline
Nancy    &  4.23\% & 29.95\% & 76.16\% & 17.05\% & 19.61\% & 53.00\% \\ \hline
Orsay    &  3.59\% & 38.98\% & 38.91\% & 27.17\% & 57.50\% & 33.85\% \\ \hline
Rennes   &  7.84\% & 35.96\% & 50.01\% & 23.38\% & 42.15\% & 40.66\% \\ \hline
Sophia   &  8.76\% & 25.64\% & 54.36\% & 19.57\% & 36.88\% & 54.79\%\\ \hline
Toulouse &  1.47\% & 28.37\% &  6.86\% & 24.85\% & 91.67\% & 46.78\% \\ \hline
\end{tabular}
\caption{Job-related statistics}
\label{jobs_stat2}
\end{table}

\item Users:

\begin{table}[H]
\centering
\begin{tabular}{|c|c|c|}
\hline
Site & Number of users & Users from other sites \\ \hline \hline
Bordeaux & 223 & 87.00\%  \\ \hline
Lille    & 198 & 65.66\%  \\ \hline
Lyon     & 169 & 73.37\%  \\ \hline
Nancy    & 157 & 83.44\%  \\ \hline
Orsay    & 172 & 76.16\%  \\ \hline
Rennes   & 213 & 82.63\%  \\ \hline
Sophia   & 187 & 85.03\%  \\ \hline
Toulouse & 160 & 81.88\%  \\ \hline
\end{tabular}
\caption{User-related statistics}
\label{users_stat}
\end{table}

\end{itemize}


\section{Conclusions}
\label{conclusion}
We have observed an activity increase for all the Grid'5000 sites compared to 2007 year~\cite{rapport2007}. The global activity of the platform has grown from 40\% to 65\%. 2007 was still a development year for Grid'5000, whereas in 2008, the platform is fully working and it reaches its cruising speed.

\listoffigures
\addcontentsline{toc}{section}{List of Figures}

\bibliographystyle{alpha}
\bibliography{references}

\newcommand{\etalchar}[1]{$^{#1}$}
\begin{thebibliography}{CCG{\etalchar{+}}05}

\bibitem[CCG{\etalchar{+}}05]{oar}
Nicolas Capit, Georges~Da Costa, Yiannis Georgiou, Guillaume Huard, Cyrille
  Martin, Gr\'egory Mouni\'e, Pierre Neyron, and Olivier Richard.
\newblock A batch scheduler with high level components.
\newblock In {\em Cluster computing and Grid 2005 (CCGrid05)}, 2005.

\bibitem[Cea05]{G5K}
F.~Cappello~et al.
\newblock Grid'5000: A large scale, reconfigurable, controlable and monitorable
  grid platform.
\newblock In {\em 6th IEEE/ACM International Workshop on Grid Computing,
  Grid'2005}, Seattle, Washington, USA, Nov. 2005.

\bibitem[IDE{\etalchar{+}}06]{IDE06}
A.~Iosup, C.~Dumitrescu, D.~Epema, Hui Li, and L.~Wolters.
\newblock How are real grids used? the analysis of four grid traces and its
  implications.
\newblock In {\em 7th IEEE/ACM International Conference on Grid Computing},
  September 2006.

\bibitem[OL09]{rapport2007}
{A}nne-{C}{\'e}cile {O}rgerie and {L}aurent {L}ef{\`e}vre.
\newblock {A} year in the life of a large-scale experimental distributed
  system: usage of the {G}rid'5000 platform in 2007.
\newblock INRIA Research Report 6965, INRIA, April 2009.

\bibitem[OLG08a]{OLG08a}
Anne-C\'ecile Orgerie, Laurent Lef\`evre, and Jean-Patrick Gelas.
\newblock Chasing gaps between bursts : Towards energy efficient large scale
  experimental grids.
\newblock In {\em PDCAT 2008 : The Ninth International Conference on Parallel
  and Distributed Computing, Applications and Technologies}, Dunedin, New
  Zealand, December 2008.

\bibitem[OLG08b]{OLG08b}
Anne-C\'ecile Orgerie, Laurent Lef\`evre, and Jean-Patrick Gelas.
\newblock Save watts in your grid: Green strategies for energy-aware framework
  in large scale distributed systems.
\newblock In {\em 14th IEEE International Conference on Parallel and
  Distributed Systems (ICPADS)}, Melbourne, Australia, December 2008.

\end{thebibliography}
\addcontentsline{toc}{section}{Bibliography}
\end{document}